\documentclass[a4paper,11pt]{article}
\usepackage{jheppub} % for details on the use of the package, please see the JINST-author-manual
\usepackage{xcolor}
\usepackage{amsmath,soul}
\usepackage{graphicx}
\usepackage{subfigure}
\def\EDIT#1{{\textcolor{red}{#1}}} 
\usepackage{hyperref}
\usepackage{caption}
\usepackage{comment}
\usepackage{subcaption}
\usepackage{booktabs}

\title{\boldmath  
Investigating (Non)-Integrability and Pulsating String in D3‐Brane Background
}

\author[a]{Rashmi R. Nayak,}
\author[b]{Kamal L. Panigrahi,}
\author[c]{Manoranjan Samal,}
\author[b]{Balbeer Singh}
\affiliation[a]{Centre for Ocean, River, Atmosphere and Land Sciences,\\
Indian Institute of Technology Kharagpur, \\
Kharagpur-721302, India}
\affiliation[b]{Department of Physics, Indian Institute of Technology Kharagpur,\\ Kharagpur 721 302, India}
\affiliation[c]{Rayagada Autonomous College, \\Rayagada, 765001, India}

\emailAdd{rashmi@coral.iitkgp.ac.in}
\emailAdd{panigrahi@phy.iitkgp.ac.in}
\emailAdd{manoranjan.phys@gmail.com}
\emailAdd{curiosity1729@kgpian.iitkgp.ac.in}

\abstract{
This work explores the (non)-integrability and chaotic dynamics of classical strings in the background of a D3-brane with a non-commutative parameter, within the framework of the AdS/CFT correspondence. Using the Polyakov action, we derive the equations of motion and constraints for pulsating strings and analyze their stability through perturbation theory. In the high-energy limit, the first-order perturbed equation simplifies to the P\"oschl-Teller equation, solvable via associated Legendre or hypergeometric functions, while numerical methods are employed for generic energy values. We demonstrate that the non-commutative parameter enhances chaotic behavior, as evidenced by the Largest Lyapunov Exponent (LLE). Furthermore, we investigate the integrability of geodesic motion and identify two distinct string modes: captured at and escape to infinity. Finally, we study pulsating strings in the deformed $(AdS_{3} \times S^{2})_{\varkappa}$ background, deriving dispersion relations for both short and long strings. 
}


 \arxivnumber{2503.02548}

\makeatletter
\gdef\@fpheader{Accepted in the EPJC}
\makeatother

\begin{document}
\maketitle
\flushbottom
% \textbf{\EDIT{The text in Red is comment/Discussion.}}
\section{Introduction}\label{sec:intro}
% \color{red} Some physical aspects of the system need to be given in the context of non-integrability and chaos. Role of non-commutative geometry.\\
% Ohta's paper\cite{Miao:2004bn} can be analysed/checked for chaos/non-integrability if needed. \\
% \color{black}
Integrability has been an interesting arena for many parts of theoretical physics. Many seemingly different fields are well-connected through the aspects of integrability \cite{beisert2012review,torrielli2016classical,zarembo2017integrability}. 
% Liouville integrability is a property of a dynamical system with n degrees of freedom, where there exist n independent conserved quantities (integrals of motion) that are in involution. This allows the system's equations of motion to be solved exactly by quadratures.
Integrable models, though rare to find, are highly significant due to their underlying symmetries, which enable exact solvability. Specifically, Liouville integrability asserts that a motion of a dynamical system is exactly solvable when all submanifolds of its phase space are integrable.
% Although integrable models are rare, their importance lies in the underlying symmetries of such models, which are therefore completely solvable.
% Specifically, Liouville integrability implies that a dynamical system is exactly solvable when its phase space is completely foliated by invariant, integrable submanifolds corresponding to conserved quantities in involution.
  \\
 In recent years, the framework of integrability has been effectively utilized within the context of AdS/CFT correspondence \cite{metsaev1998type,maldacena1999large,babichenko2010integrability,bombardelli2016integrability}. One important question is the criterion for integrability in gauge field theories. Even though matching string states and their dual operators with the one-to-one correspondence is extremely challenging, exploring classical string solutions in different $AdS\times S$ backgrounds has been crucial in advancing this correspondence. Due to the presence of integrability on both sides of the duality, it has become possible to establish a connection between the Bethe equations governing spin chains and the worldsheet symmetries in the classical $AdS_{5}\times S^{5}$ string sigma model 
 % \cite{delduc2014integrable}
 \cite{Bena:2003wd}. This proves the existence of an infinite number of conserved quantities associated with string motion in this background. However, in general, proving the integrability of the background is a cumbersome task to achieve which requires constructing the Lax connection, for which there is no standard way of deriving it. To this end, a notable approach is to use analytic non-integrability technique, as first applied in \cite{basu2011analytic,basu2011chaos}. The process involves identifying solutions for classical string motion in the gravitational background corresponding to a given gauge theory, formulating it as a Hamiltonian system, and then using the Normal Variational Equation (NVE) \cite{ruiz1999differential} and the Kovacic algorithm \cite{KOVACIC19863} to establish the non-integrability of the system. The Kovacic algorithm tests whether the differential equations of the system admit Liouville integrable solutions, which are defined as solutions that can be expressed through a finite sequence of algebraic operations and integrations (i.e., quadratures). The absence of such solutions confirms the non-integrability of the system.
 % allowing the application of variational methods to analyze (non)-integrability. 
 At the classical level, previous studies \cite{giataganas2014marginal,roychowdhury2017analytic,basu2011integrability,rigatos2020nonintegrability,Rigatos:2020igd,penin2024evidence} have demonstrated that string motion in various curved backgrounds is non-integrable, and also exhibiting signatures of chaotic dynamics \cite{frolov1999chaotic,pando2010chaos,basu2012chaos,basu2017chaotic,dutta2025chaos}. In contrast, the motion of a test particle in most of these radially symmetric backgrounds remains integrable \cite{stepanchuk2013non,Chervonyi:2013eja,giataganas2017non}, suggesting that the extended nature of a string leads to more complex behaviour. \\
% Beyond the chaotic aspects of motion, there are multiple motivations for investigating string dynamics in a black hole background...\\
% \newline
% All these studies have been done largely in commutative backgrounds and non-commutative bkg are less explored arena still...However, in this work, we aim to investigate the integrability of D3 brane with the non-commutative parameter from the perspective of dual gravity... 
% \EDIT{role of non-commutative geo.; non-commutative parameter from the T-duality}. \\
% \cite{Fischler:2018kwt}
Most of the studies have focused on backgrounds dual to commutative gauge theories, while background description of non-commutative gauge theories remain a relatively less explored domain \cite{Fischler:2018kwt}. The reason is that there are various backgrounds with a non-trivial B-field that have been explored, however, due to the form of the B-field they lead to commutative field theories. However, in this work, we seek to examine the integrability of the D3-brane with a non-commutative parameter from the dual gravity. The D3-brane with a constant NS-NS two-form field is a $D=10$-dimensional string theory solution obtained by series of T-duality operations. All solutions with nonzero B-fields preserve 16 supersymmetries and in particular for simplicity, we assume that only $B_{23}$ is non-zero then the background metric in the string frame is given by \cite{Maldacena:1999mh,Breckenridge:1996tt,Hashimoto:1999ut}
\begin{align} \label{metric}
    ds^2 &= f^{-1/2} \Big[-dx_{0}^2 + dx_{1}^2 + h(dx_{2}^2 + dx_{3}^2) \Big] + f^{1/2} (dr^2 + r^2 d\Omega_{5}^2),
\end{align}
where, 
$$f= 1+ \frac{\alpha^{\prime^{2}} R^{4}}{r^4},  \qquad h^{-1}= \sin^2 \theta f^{-1} + \cos^2 \theta,$$
and,
$$ B_{23}= \frac{\sin\theta}{\cos\theta} f^{-1} h, \quad
F_{01r}= \frac{\sin\theta ~ \partial_{r} f^{-1}}{g}, \quad
F_{0123r}= \frac{ h~\cos\theta ~ \partial_{r} f^{-1}}{g}, \quad e^{2 \phi} = g^2 h.$$
\newline
The asymptotic value of the $B$ field is $B^{\infty}_{23}=\tan \theta$. The parameter $R$ is defined as $\cos \theta R^{4}=4 \pi g N $, where $N$ is the number of D3-branes and $g\equiv g_{\infty}$ is the asymptotic value of the coupling constant. 
% When $\theta$ goes to zero, we recover the ordinary D3-brane with $AdS_{5}\times S^5$ as the near-horizon geometry. In the asymptotic limit $r \xrightarrow{} \infty$, the background geometry reduces to flat spacetime. 
In the throat region, it contains both the NS-NS and RR fields \cite{Maldacena:1999mh}. 
We have set up a system where the non-commutative geometry exists in the background, but the string dynamics, due to the choice of the ansatz$^{\S\ref{set-up}}$, do not interact. 
%This means the $B$-field does not enter the equations of motion or contribute to the action explicitly. 
%The parameter $\theta$ still characterizes the non-commutative structure of the background spacetime. Even though the string ansatz does not reveal this structure explicitly, the background remains intrinsically non-commutative. 
This can be interpreted as a manifestation of ``frozen"  non-commutative directions. So, $\theta$ acts as a background non-commutative parameter.
% Thus, we are studying closed strings in a geometry dual to a non-commutative gauge theory. 
We would like to study first the (non)-integrability in this background set-up by applying the NVE approach and the Kovacic algorithm and then check the presence of chaos in the system by numerically computing the (largest) Lyapunov exponent.\\
\newline
Within the AdS/CFT correspondence framework, the linearized perturbations of semiclassical strings are useful for extending the duality beyond the leading order \cite{larsen1994strings,khan2006improved,kiosses2014second,bhattacharya2017perturbations,bhattacharya2018perturbations}. The key motivations for exploring perturbative solutions are to examine the stability of string configurations and to compute quantum string corrections to the expectation value of Wilson loops.
Motivated by this, 
% we analytically and numerically derive and then analyze the perturbation equations for the pulsating string solution in the D3-brane. 
by using the general formalism of the construction of the perturbation equations via the Polyakov action \cite{larsen1994propagation,barik2018perturbation}, we first compute the geometric covariant quantities like normal fundamental form and extrinsic curvature tensor to write the perturbation equations. Then, using the pulsating string ansatz, we derive solutions to the equations of motion and constraints in the D3-brane without the
NS-NS flux. For high-energy \(\mathcal{E}\), the first-order perturbed equation, expressed in Fourier series form, simplifies to the Pöschl-Teller equation, whose solutions are given in terms of associated Legendre functions or hypergeometric functions. For generic values of \(\mathcal{E}\), the corresponding second-order equation is solved numerically. 
% This study is expected to aid in determining the first-order correction to the energy, which corresponds to the anomalous dimension of gauge theory operators in the strong coupling regime.
\newline
% The role of integrability in AdS/CFT duality is like a tool to provide the access to spectrum on both sides which requires to identify the classical solutions of the $AdS_{5}\times S^5$ with the global charges. Usually, the solutions are taken in some specific limit/sector of the whole string theory to make it more tractable.  In this context, a large variety of rotating and spinning strings has been studied in Ad S5 × S5 precisely and also have been mapped to dual spin-chain excitations.
\\
Integrable string backgrounds such as \( \text{AdS}_n \times S^n \), arising as near-horizon limits of brane configurations, are of particular interest.
In certain integrable string backgrounds, such as $AdS_5 \times S^5$, integrability serves as a valuable tool, enabling exact computations of the spectra on both sides of the AdS/CFT duality by matching classical solutions with their corresponding global charges
\cite{maldacena1999large,gubser1998gauge,witten1998anti}. Typically, these solutions are examined within specific limits or sectors of the full string theory to simplify the analysis \cite{berenstein2002strings,gubser2002semi,minahan2003circular}. In this setting, a wide range of rotating and spinning string configurations in \(AdS_5 \times S^5\) have been thoroughly studied and mapped to their dual spin-chain excitations \cite{frolov2002semiclassical,russo2002anomalous,tseytlin2003semiclassical,frolov2003multi,frolov2003rotating,arutyunov2003spinning}. Furthermore, these solutions have been extended to incorporate the well-established giant magnon \cite{hofman2006giant,khouchen2014giant,ahn2014finite}, folded string \cite{de1996planetoid}, and spiky string configurations \cite{kruczenski2005spiky}, with their corresponding gauge theory duals studied in depth.
Although the exact gauge theory operators corresponding to this set of string states have not yet been identified, they still remain interesting.
Equally important are related integrable deformations of the backgrounds with non-trivial parameters.
% This is a deformation of integrable sigma models (e.g., the Principal Chiral Model or AdS/CFT string sigma models) based on solutions to the Classical Yang-Baxter Equation (CYBE). It introduces a deformation parameter (often denoted as 
% η
% η) that modifies the target space geometry while preserving integrability.
The Yang Baxter deformation of \( AdS_5 \times S^5 \) provides insight into the relationship between integrability and the role of global symmetries preserved by the target spacetime \cite{klimcik2003yang,klimcik2008integrability,matsumoto2014lunin,klimcik2014integrability,arutyunov2016scale,hoare2016homogeneous,osten2017abelian,araujo2017yang}. It changes the structure of the supercoset $\frac{PSU(2,2|4)}{SO(1,4)\times SO(5)}$ by introducing a continuous parameter, commonly denoted by $\varkappa$ with $\varkappa ~\epsilon ~[0,\infty)$, which controls the strength of the deformation. This deformed background has the corresponding non-commutative gauge theory \cite{kameyama2015new,araujo2018conformal,hatsuda2023exact} through the gravity/CYBE correspondence.
% The q-deformation changes the Lie algebra of the classical charges by its quantum-deformed version,
% ...which is then incorporated into the superstring action for $AdS_5 \times S^5$ having a real deformation parameter $\chi \epsilon [0, 1)$ 
% or equivalently another parameter called $\chi$. 
% Here, we aim to investigate 
Various studies have been done in the integrable deformed  $(AdS \times S)_{\varkappa}$ backgrounds \cite{khouchen2014giant,banerjee2014rotating,panigrahi2014pulsating,arutyunov2015exact,banerjee2016circular,hernandez2017spinning,banerjee2018probing}.
Since pulsating string solutions in specific subsectors of the deformed background exhibit greater stability than the rotating strings, in this work, we examine the deformed $(AdS_{3}\times S^2)_{\varkappa}$ sector of the deformed \(( \text{AdS}_5 \times S^5)_{\varkappa} \).
% , which corresponds to a non-commutative field theory on the gauge theory side. 
We study a string pulsating in deformed \( \text{AdS}_3 \) with its centre of mass rotating on \( S^2 \) and write down the energy spin relationship for short and long strings.\\
% and analyze the solution in terms of energy as function of oscillation number for a class of pulsating strings in the deformed AdS3 \\
%In the near-horizon of D-brane bkg, we have, in the decoupling limit, where the integrability is restored... \\
% \EDIT{but why deformed one?; connect to deformed one} 
\\
The rest of the paper is organized as follows. In Section \ref{set-up}, we begin with the Polyakov action to derive the Lagrangian and obtain Hamilton’s equations of motion for the appropriate string ansatz. This section establishes the framework required for the subsequent analysis. In Section \ref{non-integ}, we demonstrate the non-integrability of the D3-brane background with the non-commutative parameter. Section \ref{particle} proves the integrability of the geodesic particle motion. Section \ref{chaos} investigates the chaotic string dynamics of the system.  In Section \ref{perturb analysis}, we perturb the string in the presence of the background and perform a stability analysis, supplemented by numerical solutions for generic parameters. Section \ref{short-long} focuses on pulsating strings in the deformed $(AdS_{3} \times S^{3})_{\varkappa}$ background, deriving dispersion relations for both short and long strings. Finally, in Section \ref{conclusion}, we summarize our findings and discuss possible future directions.
\color{black}
%References: \cite{PhysRevD.62.047503,Kaya:1999vb}
\section{Set up: Closed string in D3-brane with non-commutative parameter}\label{set-up}
The Polyakov action is given as,
\begin{equation} \label{eqn: polyakov}
 S = -\frac{1}{2\pi \alpha^{\prime}} \int d\sigma d\tau \Big( \sqrt{-g}  g^{\alpha \beta} G_{\mu \nu}\partial_{\alpha} X^{\mu}\partial_{\beta} X^{\nu} - \epsilon^{\alpha \beta} \partial_{\alpha} X^{\mu} \partial_{\beta} X^{\nu} B_{\mu \nu} \Big),
 \end{equation}
 where $\alpha^{\prime} = l_s^{2}$ ($l_s$ represents the string length). $X^{\mu}$ represents the target space co-ordinates, $G_{\mu \nu}$ is the target space metric and $g_{\alpha \beta}$ is the worldsheet metric. We utilise the reparameterization and Weyl symmetries of the action and fix the conformal gauge, $g^{\alpha\beta}$ = $\eta^{\alpha\beta}$. 
In this gauge, the vanishing of energy-momentum tensor $T_{\alpha \beta}$ = 0 leads to the Virasoro constraints
 \begin{flalign}
 G_{\mu\nu} \partial_{\tau}X^{\mu} \partial_{\sigma}X^{\nu}& = 0, \label{eqn: gauge_conformal}\\
G_{\mu\nu} \Big( \partial_{\tau}X^{\mu} \partial_{\tau}X^{\nu}+\partial_{\sigma}X^{\mu} \partial_{\sigma}X^{\nu} \Big)& = 0.
\label{vir-1}
\end{flalign}
%\color{red} Find the Lag and Hamiltonian first. Then equation of motion. \\
\color{black}
 We consider the following ansatz 
 % \footnote{One can choose more complicated ansatz, but even being a simpler ansatz, it is sufficient to establish the non-integrability in the system as we will see below. However, this ansatz does not explicitly give the appearance of the $B_{23}$-field in the Lagrangian and in the equation of motions.
 %%% , unlike, for instance, \cite{penin2024evidence}. 
 % \EDIT{In general, there is no problem with the B-field not appearing in the Lagrangian/Hamiltonian, as non-integrability is manifested from the form of the background geometry.}
 %%% Such a similar situation was observed earlier in ref \cite{penin2024evidence}
 % }
 \cite{penin2024evidence,giataganas2017non}:
\begin{align}
  t = t(\tau) \ ,  r= r(\tau) \ , x_{i}= \textit{constant} \ , i=1,2,3, \nonumber \\ 
   \phi_{1} = \phi_{1}(\tau) \ , \phi_{2} = \phi_{2}(\tau) \ , \phi_{3} = n_{3}\sigma  \ , \phi_{4} = n_{4}\sigma  \ , \phi_{5} = n_{5}\sigma.
\end{align}
% \color{red} The other choice of ansatz \cite{Banerjee:2019puc}:
 %\begin{eqnarray}
%     x_{i}= \alpha_{i} f_{i}(\sigma) + \beta_{i} g_{i}(\tau),~~i=2,3 
% \end{eqnarray}
 \color{black}
The $n_{j}$'s ($j=3,4,5$) represent the number of times the string is wrapped around the angle coordinate $\phi_{3}, \phi_{4}, \phi_{5}$ respectively.
%\color{red} In this, what metric of $d\Omega_{5}^2$ is to be chosen?? Hopf coordinates? or not Hopf? or Dimitris' way?\\
%\color{black}
The $S^5$ metric is given by $d\Omega_{5}^2= d\phi_{1}^2 + \cos^2 \phi_{1} d\phi_{2}^2 + \sin^2 \phi_{1} \Big(d\phi_{3}^2 + \cos^2\phi_{3} d\phi_{4}^2 + \sin^2 \phi_{3} d\phi_{5}^2\Big)$. Then the Lagrangian is given by
 % \begin{align}
 %    \mathcal{L} &= -\frac{1}{2 \pi \alpha \prime} \Bigg[ \frac{1}{\sqrt{f}} \dot t^2 - \sqrt{f} \dot r^2 +\sqrt{f} r^2 \Big(-\dot \phi_{1}^2 -  \cos^2 \phi_{1} \dot \phi_{2}^2 + \sin^2 \phi_{1} n_{3}^2 + \sin^2 \phi_{1} \cos^2 \phi_{3} n_{4}^2 \nonumber \\
 %    &\quad  + \sin^2 \phi_{1} \sin^2 \phi_{2} n_{5}^2\Big) \Bigg]
 % \end{align}
 \begin{align}
     % \mathcal{L} &= -\frac{1}{2 \pi \alpha \prime} \Bigg[ \frac{1}{\sqrt{f}} \dot t^2 - \sqrt{f} \dot r^2 +\sqrt{f} r^2 \Big(-\dot \phi_{1}^2 -  \cos^2 \phi_{1} \dot \phi_{2}^2 + \sin^2 \phi_{1} n_{3}^2 + \sin^2 \phi_{1} \cos^2 \phi_{3} n_{4}^2 \nonumber \\
     % &\quad  + \sin^2 \phi_{1} \sin^2 \phi_{3} n_{5}^2\Big) \Bigg]. \\
    \mathcal{L} &= -\frac{1}{2\pi\,\alpha'}\Biggl[\frac{1}{\sqrt{f}}\,\dot t^2
           \;-\;\sqrt{f}\,\dot r^2 +\;\sqrt{f}\,r^2\Bigl(
              -\,\dot\phi_{1}^2
           \;-\;\cos^2\!\phi_{1}\,\dot\phi_{2}^2
           \;+\;\sin^2\!\phi_{1}\,n_{3}^2 \nonumber\\
           &\quad 
           \;+\;\sin^2\!\phi_{1}\,\cos^2\!\bigl(n_{3}\sigma\bigr)\,n_{4}^2 
      +\;\sin^2\!\phi_{1}\,\sin^2\!\bigl(n_{3}\sigma\bigr)\,n_{5}^2
    \Bigr)
\Biggr]\,,
 \end{align}
where the dot above a coordinate denotes differentiation with respect to the time on the world sheet, $\tau$ and that a prime denotes differentiation with respect to $\sigma$.
 Note that the presence of $\theta$ in $f$ through $R$ makes the above analysis suited for the non-commutative case. Next, the Hamiltonian is obtained as
\begin{align}
    % \mathcal{H} &= \frac{1}{2} \sqrt{f} n_5^2 r^2 \sin ^2(\text{$\phi_1$}) \sin ^2(\text{$\phi_3$})+\frac{1}{2} \sqrt{f} n_3^2 r^2 \sin ^2(\text{$\phi_1$})+\frac{1}{2} \sqrt{f} n_4^2 r^2 \sin ^2(\text{$\phi_1$}) \cos ^2(\text{$\phi_3$}) \nonumber \\
    % &\quad +\frac{p_{\text{$\phi_2$}}^2 \sec ^2(\text{$\phi_1$})}{2 \sqrt{f} r^2}+\frac{p_{\text{$\phi_1$}}^2}{2 \sqrt{f} r^2}+\frac{p_r^2}{2 \sqrt{f}}-\frac{1}{2} \sqrt{f} p_t^2.
    \mathcal{H}
&= \tfrac12\,\sqrt{f}\,n_{5}^{2}\,r^{2}\,\sin^{2}\!\phi_{1}\,\sin^{2}\!\bigl(n_{3}\sigma\bigr)
   + \tfrac12\,\sqrt{f}\,n_{3}^{2}\,r^{2}\,\sin^{2}\!\phi_{1}
   + \tfrac12\,\sqrt{f}\,n_{4}^{2}\,r^{2}\,\sin^{2}\!\phi_{1}\,\cos^{2}\!\bigl(n_{3}\sigma\bigr) \nonumber\\
&\quad
   + \frac{p_{\phi_{2}}^{2}\,\sec^{2}\!\phi_{1}}{2\,\sqrt{f}\,r^{2}}
   + \frac{p_{\phi_{1}}^{2}}{2\,\sqrt{f}\,r^{2}}
   + \frac{p_{r}^{2}}{2\,\sqrt{f}}
   - \tfrac12\,\sqrt{f}\,p_{t}^{2}\,.
\end{align}
As a consistency check, it is easy to verify that the non-trivial Virasoro condition eq (\ref{vir-1}) provides the same equation $\mathcal{H}=0$.\\
Therefore, the equations of motion are obtained as follows:
\begin{align}
    \dot t &= -\sqrt{f} p_{t}, \\
    \dot r &= \frac{p_{r}}{\sqrt{f}}, \\
    \dot \phi_{1} &= \frac{p_{\phi_{1}}}{\sqrt{f} r^2}, \\
    \dot \phi_{2} &= \frac{p_{\phi_{2}} \sec ^2 \phi_{1}}{\sqrt{f} r^2}, \\
    %% \dot \phi_{3} = 0, ~~\dot \phi_{4}=0,~~\dot \phi_{4}=0
    %\dot p_{r} &= -\sqrt{f} n_5^2 r \sin ^2\text{$\phi_1$} \sin ^2\text{$\phi_3$}-\sqrt{f} n_3^2 r \sin ^2\text{$\phi_1$}-\sqrt{f} n_4^2 r \sin ^2\text{$\phi_1$} \cos ^2\text{$\phi_3$} \nonumber \\
    %&\quad  + \frac{p_{\text{$\phi_2$}}^2 \sec ^2\text{$\phi_1$}}{\sqrt{f} r^3}+\frac{p_{\text{$\phi _1$}}^2}{\sqrt{f} r^3}, \\
     \dot p_{r}
        &= -\,\sqrt{f}\,n_{5}^{2}\,r\,\sin^{2}\!\phi_{1}\,\sin^{2}\!\bigl(n_{3}\sigma\bigr)
        -\,\sqrt{f}\,n_{3}^{2}\,r\,\sin^{2}\!\phi_{1}
        -\,\sqrt{f}\,n_{4}^{2}\,r\,\sin^{2}\!\phi_{1}\,\cos^{2}\!\bigl(n_{3}\sigma\bigr) \\
        &\quad
        + \frac{p_{\phi_{2}}^{2}\,\sec^{2}\!\phi_{1}}{\sqrt{f}\,r^{3}}
      + \frac{p_{\phi_{1}}^{2}}{\sqrt{f}\,r^{3}}\,,\\
    %\dot p_{\phi_1} &= -\sqrt{f} n_5^2 r^2 \sin \text{$\phi _1$} \cos \text{$\phi _1$} \sin ^2\text{$\phi_3$}-\sqrt{f} n_4^2 r^2 \sin \text{$\phi _1$} \cos ^3\text{$\phi_3$} \cos \phi_{1}\nonumber \\
    %&\quad 
        %% +\sqrt{f} n_4^2 r^2 \sin ^3\text{$\phi $1} \cos \text{$\phi $1}
    %-\sqrt{f} n_3^2 r^2 \sin \text{$\phi_1$} \cos \text{$\phi_ 1$}-\frac{p_{\text{$\phi _2$}}^2 \tan \text{$\phi _1$} \sec ^2\text{$\phi _1$}}{\sqrt{f} r^2},\\
     \dot p_{\phi_{1}}
 &= -\,\sqrt{f}\,n_{5}^{2}\,r^{2}\,\sin\!\phi_{1}\,\cos\!\phi_{1}\,\sin^{2}\!\bigl(n_{3}\sigma\bigr)
  -\,\sqrt{f}\,n_{4}^{2}\,r^{2}\,\sin\!\phi_{1}\,\cos\!\phi_{1}\,\cos^{3}\!\bigl(n_{3}\sigma\bigr)
 \nonumber\\
 &\quad -\,\sqrt{f}\,n_{3}^{2}\,r^{2}\,\sin\!\phi_{1}\,\cos\!\phi_{1}
    -\,\frac{p_{\phi_{2}}^{2}\,\tan\!\phi_{1}\,\sec^{2}\!\phi_{1}}{\sqrt{f}\,r^{2}}\,,\\
    \dot p_{\phi_2} &= 0,\\
     \dot p_{\phi_3} &= 0,\\
     % \sqrt{f} n_{4}^2 r^2 \cos\phi_{3} \sin^2\phi_{1} \sin\phi_{3}-\sqrt{f} n_5^2 r^2 \sin ^2\text{$\phi_1$} \sin \text{$\phi _3$} \cos \text{$\phi_3$},\\
     \dot p_{\phi_4} &= 0,\\
     \dot p_{\phi_5} &= 0.
\end{align} 
In the next section, we establish the non-integrability in the closed string hovering near the D3-brane background with the non-commutative parameter.
% \newline
% \color{red} 1. Next, find out the the invariant plane and then the NVE. Then finally checking chaos numerically. \\
% 2. Dispersion relations (check if any done). 
% \color{black} 
\section{Classical string solution and (non)-integrability}\label{non-integ}
In this section, we shall examine the non-integrability of the closed string in the above background using the Kovacic algorithm. The Lagrange's equations of motion are obtained as
\begin{align}
\dot{t} & = E \sqrt{f}, \label{eq-t} 
\end{align}
\begin{align}
\ddot{r} & + \frac{f' \dot{r}^2}{4 f} - \frac{f' \dot{t}^2}{4 f^2} - \frac{r^2 f' \dot{\phi}_{1}^2}{4 f} - \cos^2\phi_{1}\, r \dot{\phi}_{2}^2 - \frac{r^2 \cos^2\phi_{1} f' \dot{\phi}_{1}^2}{4 f} - r \dot{\phi}_{1}^2 \nonumber\\[1mm]
& \quad + \frac{n_{3}^2 r^2 \sin^2 \phi_{1} f'}{4 f} + n_{3}^2 r \sin^2 \phi_{1} + \frac{n_{4}^2 \cos^2\phi_{3}\, r^2 \sin^2\phi_{1} f'}{4 \sqrt{f}} + n_{4}^2 \cos^2\phi_{3}\, r \sin^2\phi_{1} \nonumber\\[1mm]
& \quad +  n_{5}^2 r \sin^2 \phi_{1} \sin^2 \phi_{3} + \frac{n_{5}^2 r^2 \sin^2 \phi_{1} \sin^2\phi_{3} f'}{4 f} = 0, 
\end{align}
\begin{align}
\ddot{\phi}_{1} & + \frac{f' \dot{r} \dot{\phi}_{1}}{2 f} + \frac{2 \dot{r} \dot{\phi}_{1}}{r} + \frac{1}{2} \Bigg( \sin 2\phi_{1}\, \dot{\phi}_{2}^2 + n_{3}^2 \sin 2 \phi_{1} + n_{4}^2 \cos^2 \phi_{3}\, \sin 2 \phi_{1} + n_{5}^2 \sin^2 \phi_{3}\, \sin 2 \phi_{1} \Bigg) = 0,
\end{align}
%%\begin{align}
%%\sqrt{f}\, r^2 \sin^2 \phi_{1} \Bigg(2 n_{4}^2 \cos\phi_{3} (- \sin \phi_{3}) + 2 n_{5}^2 \sin \phi_{3} \cos \phi_{3} \Bigg) & = 0, \label{eq-phi}
%%\end{align}
% \color{red}
\begin{align}
% f^{1/2} r^2 \cos^2 \phi_1 \ddot{\phi}_2 + 2 f^{1/2} r \dot{r} \cos^2 \phi_1 \dot{\phi}_2 + \frac{1}{2} f^{-1/2} f' \dot{r} r^2 \cos^2 \phi_1 \dot{\phi}_2 - 2 f^{1/2} r^2 \cos \phi_1 \sin \phi_1 \dot{\phi}_1 \dot{\phi}_2 = 0,
\ddot{\phi}_{2} & + \frac{f' \dot{r} \dot{\phi}_{2}}{2 f} + \frac{2 \dot{r} \dot{\phi}_{2}}{r} - \frac{2\sin \phi_{1}}{\cos\phi_{1}}\, \dot \phi_{1}\dot{\phi}_{2} = 0,  \label{eq-phi}
\end{align}
\color{black}
% \begin{align} 
%     \dot t &= E \sqrt{f} \label{eq-t} \\
%     \ddot r + \frac{f'(r) \dot r^2}{4 f} - \frac{f'(r) \dot t^2}{4 f^2} - \frac{r^2 f' \dot \phi_{1}^2}{4 f} - \cos^2\phi_{1} r \dot \phi_{2}^2 - \frac{r^2 \cos^2\phi_{1} f' \dot \phi_{1}^2}{4 f} - r \dot \phi_{1}^2 \nonumber \\ 
%     &\quad + \frac{n_{3}^2 r^2 \sin^2 \phi_{1} f'}{4 f} + n_{3}^2 r \sin^2 \phi_{1} + \frac{n_{4}^2 \cos^2\phi_{3} r^2 \sin^2\phi_{1} f'}{4 \sqrt{f}} + n_{4}^2 \cos^2 \phi_{3}^2 r \sin^2\phi_{1} \nonumber \\
%     &\quad +  n_{5}^2 r \sin^2 \phi_{1} \sin^2 \phi_{3} + \frac{n_{5}^2 r^2 \sin^2 \phi_{1} \sin^2\phi_{3} f'}{4 f} =0 \\
%     \ddot \phi_{1} + \frac{ f' \dot r \dot \phi_{1}}{2 f} + \frac{2 \dot r \dot \phi_{1}}{r} + \frac{1}{2} \Bigg( \sin 2\phi_{1} \dot \phi_{2}^2 + n_{3}^2 \sin 2 \phi_{1} + n_{4}^2 \cos^2 \phi_{3} \sin 2 \phi_{1} + n_{5}^2 \sin^2 \phi_{3} \sin 2 \phi_{1} \Bigg) &=0 \\
%     \sqrt{f} r^2 \sin^2 \phi_{1} \Bigg(2 n_{4}^2  \cos\phi_{3} (- \sin \phi_{3}) + 2 n_{5}^2  \sin \phi_{3} \cos \phi_{3} \Bigg) &=0 \label{eq-phi}
% \end{align}
while the equations of motions for $\phi_{3}, \phi_{4}$ and $\phi_{5}$ are trivial. Note that here the dot represents the derivative with respect to $\tau$, but the prime denotes the derivative with respect to $r$. Let us consider an invariant plane 
$$r= \overline{r} (\tau), \quad p_{r}=\overline{r} \sqrt{f(\overline{r})}, \quad \phi_{1}= \frac{\pi}{2}, \quad p_{\phi_{1}}=0,$$ while we fix $\phi_{2}=  \phi_{3}= \frac{\pi}{2}$ and $n_{4}=0$. Then from equation \ref{vir-1}, we obtain the desired solution $\overline{r}$:
\begin{eqnarray}
    \overline{r}(\tau)= \frac{E}{\sqrt{n_{3}^2 + n_{5}^2}} \sin\left(\sqrt{n_{3}^2 + n_{5}^2} \tau\right),
\end{eqnarray}
where, we have taken $r(0)=0$.
One can verify that the above plane is indeed a solution satisfying the equations \ref{eq-t}-\ref{eq-phi}. Next, we expand the equation of motion for $\phi_{1}$ by using $\phi_{1}(\tau)= \frac{\pi}{2} + \xi(\tau)$ ; $|\xi(\tau)|<<1$. Then the corresponding NVE up to the leading order in $\xi$ is given by
\begin{equation}
      % \xi^{''}(\tau) + \Bigg(  \frac{ f^{'}(\overline{r}) \dot{\overline{r}}}{2 f(\overline{r})}  + \frac{2 \dot{\overline{r}}}{\overline{r}}\Bigg) \xi^{'}(\tau) - (n_{3}^{2} + n_{5}^2) \xi(\tau) =0,
      \frac{d^2\xi}{d\tau^2} + \Bigg(  \frac{ f^{'}(\overline{r}) \dot{\overline{r}}}{2 f(\overline{r})}  + \frac{2 \dot{\overline{r}}}{\overline{r}}\Bigg) \frac{d\xi}{d\tau} - (n_{3}^{2} + n_{5}^2) \xi(\tau) =0,
\end{equation}
which is not in the desired form of the algebraic coefficients. So, we make the substitution $\tau \xrightarrow{} x= sin\sqrt{n_{3}^2 + n_{5}^2} \tau$, which gives
\begin{eqnarray}
    % \xi^{''} + \Bigg[ \frac{2 E}{\sqrt{n_{3}^2 + n_{5}^2} x} - \frac{2 (n_{3}^{2} + n_{5}^{2})^2 R^4 \alpha{'}}{x (x^4 E^4 + (n_{3}^2 + n_{5}^2) R^4)} -\frac{x}{1- x^2}\Bigg] \xi^{'} - \frac{\xi}{1-x^2}=0.
    \frac{d^2\xi(x)}{dx^2} + \Bigg[ \frac{2 E}{\sqrt{n_{3}^2 + n_{5}^2} x} - \frac{2 (n_{3}^{2} + n_{5}^{2})^2 R^4 \alpha{'}}{x (x^4 E^4 + (n_{3}^2 + n_{5}^2) R^4)} -\frac{x}{1- x^2}\Bigg] \frac{d\xi}{dx}- \frac{\xi(x)}{1-x^2}=0.
\end{eqnarray}
With the help of the Kovacic algorithm\footnote{We have employed the Maple in-built function kovacicsols for checking of the Liouvillian solution.}, it can be shown that the above NVE does not possess the Liouvillian solution. 
%\color{red}..Some Galois theory arguments..
\color{black} 
%One can do a more detailed analysis of the NVE by changing it to the normal or Schrodinger form (more details given in the appendix).
Thus, the closed string in the presence of D3-brane having the non-commutative parameter is non-integrable. 
% \EDIT{Moreover, one can see that even switching off the non-commutative parameter, the non-integrability is persistent.} 

Here, during the analysis, we have assumed $n_{3} \ne 0$ and $n_{5} \ne 0$ throughout.
% ; otherwise.
% ,  the string would shrink to the point-particle case, which has been analysed in section \ref{particle}.
In the next section, we deal with the special case where $n_{3}=n_{4}=n_{5}=0$, which consequently shrinks the string to a point particle scenario (or geodesic).
% \begin{eqnarray}
% \xi''(x) + \Bigg[ \frac{2E}{\sqrt{n_3^2+n_5^2}\,x} - \frac{2 (n_3^2+n_5^2)^2\,\left(\frac{4\pi gN}{\cos\theta}\right)\,\alpha'}{x\Bigl(x^4E^4+(n_3^2+n_5^2)\,\left(\frac{4\pi gN}{\cos\theta}\right)\Bigr)} - \frac{x}{1-x^2} \Bigg] \xi'(x) - \frac{\xi(x)}{1-x^2} = 0.
% \end{eqnarray}
% Since the study of the integrability in the classical Hamiltonian systems is related to the variations of the trajectories of the phase space of the system, therefore in the next section, we numerically verify the non-integrability of the circular string by investigating the chaotic motion in the phase space curves. 
% \newline
% \section{Factorisation scattering}
\section{Integrability of geodesic motion} \label{particle}
%\color{red}I shall largely follow the ref \cite{Stepanchuk:2012xi}.
\color{black}
 In this section, we discuss the point-particle scenario. Since the NVE and the Kovacic algorithm provide necessary but not sufficient conditions for integrability \cite{KOVACIC19863}, we demonstrate integrability by showing that the number of integrals of motion equals the number of degrees of freedom. This can be done by assuming $n_{3}=n_{4}=n_{5}=0$ i.e., the string is not allowed to expand and is thus localised in the subspace of $S^{5}$. However, as noted in \cite{stepanchuk2013non}, the B-field does not couple to the point-like string. 
  % The metric \ref{metric} for $p=3$ possesses the $(p+1)$ symmetries along the coordinates $x^{\mu}$, $\mu=0,1,...,p$ and so provides the $(p+1)$ integrals of motions. Then, by parametrizing the sphere embedded into one higher-dimensional sphere one can show that $D-p-2$ constants of motion exist for $(D-p-2)$-sphere \cite{stepanchuk2013non}. Thus the total integrals of motion become equal to $D$, which is the total number of available degrees of freedom.\\
  The metric \ref{metric} for $D3$-brane possesses the four symmetries along the coordinates $x^{\mu}$, $\mu=0,1,2,3$ and so provides the four integrals of motions. Then, by parametrizing the sphere embedded into one higher-dimensional sphere, one can show that five constants of motion exist for $S^5$ \cite{stepanchuk2013non}. One additional constant of motion is provided by Hamiltonian, thus the total integrals of motion become equal to $4+5+1=10$, which is the total number of available degrees of freedom.\\
Furthermore, we can directly analyse the dynamics by solving the equations of motion, subject to the Virasoro constraint equation \ref{vir-1} (other Virasoro constraint equation \ref{eqn: gauge_conformal} is trivially satisfied), for the point-particle scenario. These are given by: 
\begin{align}
    r:~~~~~~~~~~\ddot{r} & + \frac{f'(r) \dot{r}^2}{4 f} - \frac{f'(r) \dot{t}^2}{4 f^2} - \frac{r^2 f' \dot{\phi}_{1}^2}{4 f} - \cos^2\phi_{1}\, r \dot{\phi}_{2}^2 - \frac{r^2 \cos^2\phi_{1} f' \dot{\phi}_{1}^2}{4 f} - r \dot{\phi}_{1}^2
\end{align}
\begin{align}
\phi_{1}:~~~~~~~~\ddot{\phi}_{1} & + \frac{f' \dot{r} \dot{\phi}_{1}}{2 f} + \frac{2 \dot{r} \dot{\phi}_{1}}{r} + \frac{1}{2} \sin 2\phi_{1}\, \dot{\phi}_{2}^2 = 0,
\end{align}
\begin{align}
 \phi_{2}: ~~~~~~~~  \ddot{\phi}_{2} & + \frac{f' \dot{r} \dot{\phi}_{2}}{2 f} + \frac{2 \dot{r} \dot{\phi}_{2}}{r} - \frac{2\sin \phi_{1}}{\cos\phi_{1}}\, \dot \phi_{1}\dot{\phi}_{2} = 0,
\end{align}
\begin{align}
 \text{Virasoro constraint}:~~~~~~~~~   \frac{-E^2 f}{\sqrt{f}} + \sqrt{f} \dot r^2 + \sqrt{f} r^2 \dot \phi_{1}^2 + \sqrt{f} r^2 \cos^2\phi_{1} \dot \phi_{2}^2=0.
\end{align}
 Solving these equations, we obtain:
\begin{align}
    r=r_{0}+ E\tau, \,\,\,\, \phi_{1}=\text{constant},\,\,\,\,\phi_{2}=\text{constant},
\end{align}
which describes a radial null trajectory with fixed angular coordinates.\\

 % \EDIT{While performing the numerical analysis, we can show the Lyapunov exponent for $n=0$ case, strengthening the above integrability argument for the geodesic case $(n=0)$. }
 % \newline
 \section{Chaos in D3 brane with non-commutative parameter} \label{chaos}
%\EDIT{In this paper \cite{fischler2018chaos}, they have found the Lyapunov exponent of the quantum chaos in the non-commutative case to be the same as that of the commutative case ie ordinary SYM theory. This is also indicative in the classical string case, as I get similar results irrespective of the parameter $\theta$ ie $\theta=0$(ordinary SYM) and $\theta \ne 0$(non-commutative SYM). }\\
%\newline
%\EDIT{We can draw plot for $\lambda$ vs $\theta$(or a) and compare!!}\\
%\newline
Since the study of the integrability in the classical Hamiltonian systems is related to the variations of the trajectories in the phase-space of the system, therefore in this section, we numerically verify the non-integrability of the circular string by investigating signatures of chaotic motion in the phase-space curves. 
We analyse the presence of chaos when the closed string comes near the D3 brane with the effect of the non-commutative parameter. 
The metric eq \ref{metric} is 
\begin{align}
    ds^2 &= f^{-1/2} \Big[-dx_{0}^2 + dx_{1}^2 + h(dx_{2}^2 + dx_{3}^2) \Big] + f^{1/2} (dr^2 + r^2 d\Omega_{5}^2),
\end{align}
where, $d\Omega_{5}^2= d\phi_{1}^2 + \cos^2 \phi_{1} d\phi_{2}^2 + \sin^2 \phi_{1} \Big(d\phi_{3}^2 + \cos^2\phi_{3} d\phi_{4}^2 + \sin^2 \phi_{3} d\phi_{5}^2\Big)$.\\
% \newline
On taking the appropriate ansatz: 
%\EDIT{old}
\begin{align}
     t = t(\tau) \ ,  r= r(\tau) \ , x_{i}= \textit{constant} \ , i=1,2,3, \nonumber \\ 
   \phi_{1} = \phi_{1}(\tau) \ , \phi_{2} = \pi/2 \ , \phi_{3} = \pi/2  \ , \phi_{4} = n_{4}\sigma  \ , \phi_{5} = n_{5}\sigma.
\end{align}
The corresponding Lagrangian is given by 
\begin{align}
    \mathcal{L}= - \frac{1}{2 \pi \alpha^{'}} \left[ \frac{\dot{t}^2}{\sqrt{f}} - \sqrt{f} \dot{r}^2 + \sqrt{f} r^2 \Big(- \dot{\phi_{1}^2} + n_{5}^2 \sin^2\phi_{1} \Big)\right].
\end{align}
The associated Hamiltonian is 
\begin{align}
    H= -\frac{1}{2} \pi  \alpha'  E^2 \sqrt{f}+\frac{\sqrt{f} n_5^2 r^2 \sin ^2\phi _1}{2 \pi  \alpha' }+\frac{\pi  \alpha'  p_{\phi _1}^2}{2 \sqrt{f} r^2}+\frac{\pi  \alpha'  p_r^2}{2 \sqrt{f}},
\end{align}
and the equations of motion are given by
\begin{align}
r' & = \frac{\pi \alpha' p_r}{\sqrt{f}},\\[1mm]
\phi'_1 & = \frac{\pi \alpha' p_{\phi_1}}{\sqrt{f} \, r^2},\\[1mm]
p_r' & = -\frac{n_5^2 r^2 f' \sin^2(\phi_1)}{4 \pi \alpha' \sqrt{f}} 
      + \frac{\pi \alpha' E^2 f'(r)}{4 \sqrt{f}}
      + \frac{\pi \alpha' f' p_{\phi_1}^2}{4 f^{3/2} \, r^2}
      + \frac{\pi \alpha' f' p_r^2}{4 f^{3/2}} \nonumber\\[1mm]
    & \quad -\frac{\sqrt{f} n_5^2 r \sin^2(\phi_1)}{\pi \alpha'}
      + \frac{\pi \alpha' p_{\phi_1}^2}{\sqrt{f} \, r^3},\\[1mm]
p_{\phi_1}' & = -\frac{\sqrt{f} n_5^2 r^2 \sin(\phi_1) \cos(\phi_1)}{\pi \alpha'}.
\end{align}
% \begin{align}
%     r'=\frac{\pi  \alpha  p_r}{\sqrt{f}}\\
%     \text{$\phi $1}'=\frac{\pi  \alpha  p_{\phi _1}}{\sqrt{f} r^2}\\
%     p_r'=-\frac{n_5^2 r^2 f' \sin ^2\left(\phi _1\right)}{4 \pi  \alpha  \sqrt{f}}+\frac{\pi  \alpha  E^2 f'(r)}{4 \sqrt{f}}+\frac{\pi  \alpha  f' p_{\phi _1}^2}{4 f^{3/2} r^2}+\frac{\pi  \alpha  f' p_r^2}{4 f^{3/2}}-\frac{\sqrt{f} n_5^2 r \sin ^2\left(\phi _1\right)}{\pi  \alpha }+\frac{\pi  \alpha  p_{\phi _1}^2}{\sqrt{f} r^3}\\
%     p_{\phi _1}'=-\frac{\sqrt{f} n_5^2 r^2 \sin \left(\phi _1\right) \cos \left(\phi _1\right)}{\pi  \alpha }
% \end{align}
The non-trivial Virasoro constraint gives rise to the constraint $H=0$ or equivalently 
% \begin{align}
%     f n_5^2 r^4 \sin ^2(\text{$\phi $1})+\pi ^2 \alpha' ^2 \left(p_{\phi _1}^2+\text{p}_{r}^2 r^2\right)=\pi ^2 \alpha'^2 E^2 f r^2.
% \end{align}
\begin{align}
    f n_5^2 r^4 \sin ^2\left(\phi _1\right)+\pi ^2 \alpha'^{2} \left(r^2 p_r^2+p_{\phi _1}^2\right)=\pi ^2 \alpha'^{2} E^2 f r^2
\end{align}
In the next subsection, we solve numerically these equations of motion and probe the presence of chaos using the Largest Lyapunov   Exponent(LLE) as a chaotic indicator. 
\subsection{(Largest) Lyapunov Exponent} 
% \EDIT{Def of LLE, size font in Figures}\\
% A hallmark of chaotic systems is that the behavior of the solution is extremely sensitive to initial conditions. The convergence of trajectories at late times implies the presence of a stable fixed point (attractor) in the system, whereas their divergence indicates an unstable fixed point (repellor). Consequently, it is essential to examine the mean exponential rate at which the distance between two neighboring trajectories evolves as a function of proper time, which is given by the following expression
% \begin{align}
% \end{align}
For a dynamical system, consider two points in phase space that are initially separated by a small distance $\delta \mathbf{x}_0$. As the system evolves, the separation between these points typically grows or shrinks exponentially over time. The Lyapunov exponent $\lambda$ quantifies this exponential rate of separation:
\begin{align}
|\delta \mathbf{x}(t)| \approx |\delta \mathbf{x}_0| e^{\lambda t},
\end{align}
A positive $\lambda$ implies exponential divergence and is indicative of chaos, while a negative value indicates that trajectories converge toward an attractor. In a dynamical system with $n$ dimensions, there are typically $n$ Lyapunov exponents, each corresponding to a different direction in phase space. The Largest Lyapunov Exponent (LLE) is the maximum of these exponents and describes the most rapid rate of separation (or convergence) of infinitesimally close trajectories.
% The largest Lyapunov exponent (LLE), often denoted by $\lambda$, quantifies the mean exponential rate at which two infinitesimally close trajectories in phase space diverge(or converge) over time. If we denote by $\delta x(0)$ the initial infinitesimal distance between two trajectories and by $\delta x(t)$ their distance after a time $t$ (or proper time), then the largest Lyapunov exponent is defined as
% \begin{align}
% \lambda = \lim_{t\to\infty} \frac{1}{t}\ln \frac{\|\delta x(t)\|}{\|\delta x(0)\|}.
% \end{align}

\begin{figure}[ht!]
    \hspace{3mm}
    \begin{subfigure}
        \centering
        \includegraphics[width=0.41\textwidth]{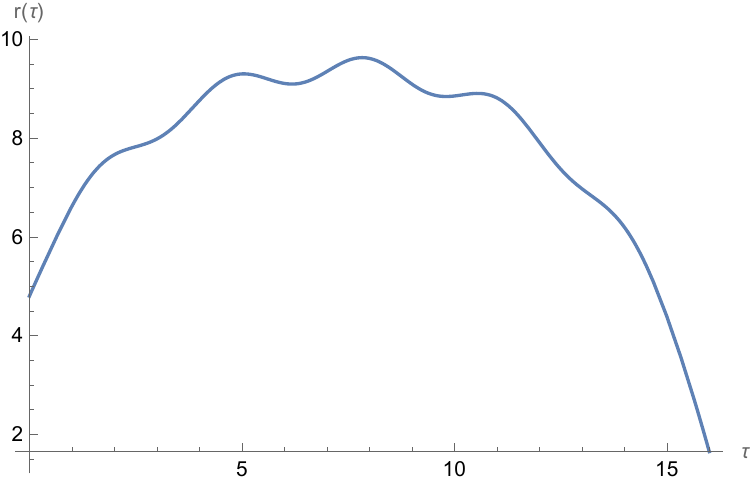}
        \put(-180,120){(a)}
    \end{subfigure}
    \hspace{12mm}
    \begin{subfigure}
         \centering
        \includegraphics[width=0.42\textwidth]{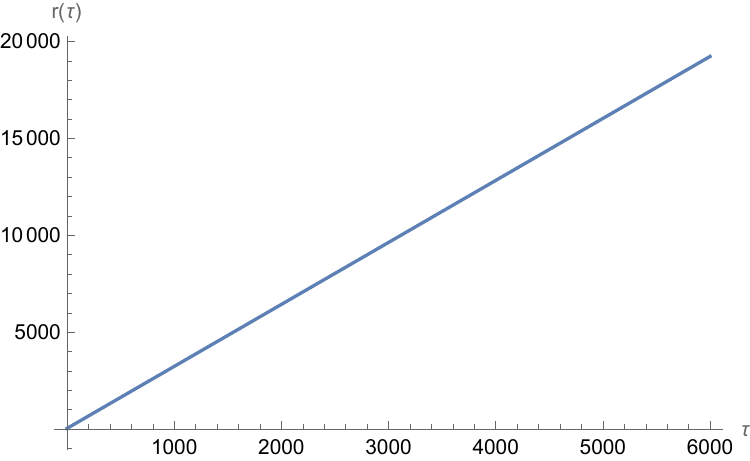}
        \put(-180,120){(b)}
    \end{subfigure}
    \caption{In (a), the string oscillates in the beginning and then gets captured into the D3 brane with initial location $r(0)=6.5$. In (b), the string escapes to infinity with the initial location $r(0)=30$. Other parameters and initial conditions are:
    $\theta=1.5,~p_{r}(0)=0.5,~\phi_{1}(0)=0,~E=1.2$.}
    \label{capture-escape}
\end{figure}

\begin{figure}[ht!]
\begin{subfigure}
    \centering
    \includegraphics[width=0.5\linewidth]{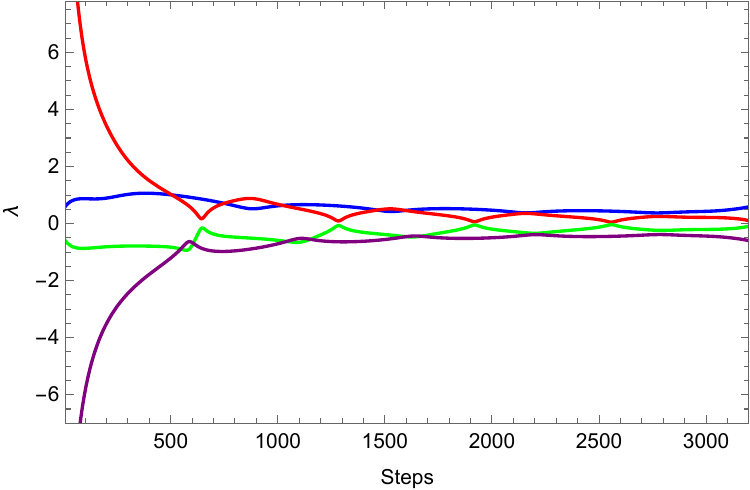}
    \put(-210,150){(a)}
    %\captionof{figure}{LLE for $\theta=0$} 
    \label{theta-0}
\end{subfigure}
\begin{subfigure}
    \centering
    \includegraphics[width=0.5\linewidth]{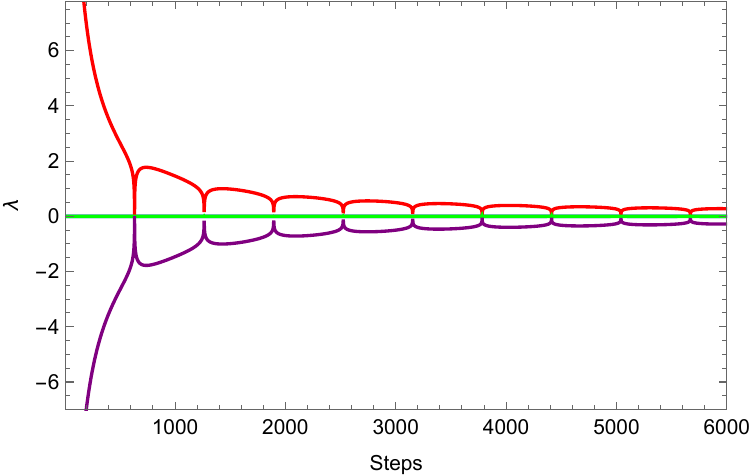}
    \put(-210,150){(b)}
    %\caption{LLE for (non-zero) $\theta=1.14$}
    % \label{LE}
\end{subfigure}
\caption{(a) shows the Lyapunov values for the capture case and  (b) shows Lyapunov values for the escape to infinity. The parameter choice and the initial condition are the same as in Fig \ref{capture-escape}.  }
\label{LE}
\end{figure}

\begin{figure}[ht!]
\begin{subfigure}
    \centering
    \includegraphics[width=0.5\linewidth]{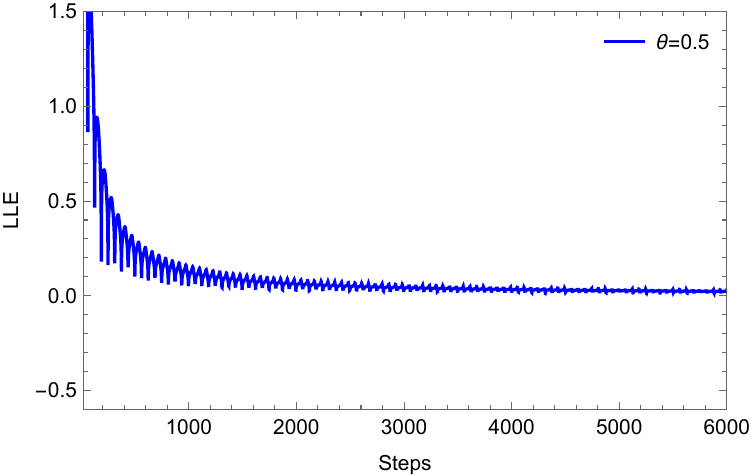}
    \put(-210,150){(a)}
    %\captionof{figure}{LLE for $\theta=0$} 
    % \label{theta-0}
\end{subfigure}
\begin{subfigure}
    \centering
    \includegraphics[width=0.5\linewidth]{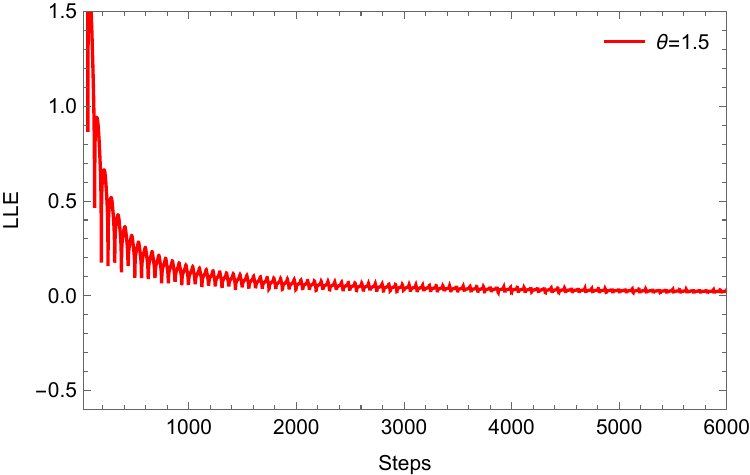}
    \put(-210,150){(b)}
    %\caption{LLE for (non-zero) $\theta=1.14$}
    % \label{LE}
\end{subfigure}
  \caption{A comparison of the Largest Lyapunov Exponent (LLE) for two scenarios: for $\theta=0.5$, the LLE is $0.037918$, while for $\theta=1.5$, it slightly increases to $0.0308643$. The initial conditions are: $r(0)=30$, $p_{r}(0)=1.2$, $\phi_{1}(0)=0$, and $E=3.2$.}
\label{LLE-comparison}
\end{figure}

In this section, we numerically check the chaos by calculating the associated (Largest) Lyapunov exponent values. Throughout the numerical calculations, without loss of generality, we have set $\alpha'=1/\pi,~ g=0.1,~ N= 500$ and $n_{5}=1$. The energy is held constant while $\theta$ is varied as a parameter. Depending on the initial location of the string we observe two kinds of string modes: captured at and escape to infinity. The capture scenario is usually observed when the string is placed near the horizon whereas the escape to infinity is observed for the string located at away from the horizon. In Fig \ref{capture-escape}, for instance, we first show the possible string trajectories and then plot the corresponding Lyapunov exponent values in Fig \ref{LE}.
% \begin{figure}[ht!]
%     \centering
%     \includegraphics[width=0.5\linewidth]{LC.pdf}
%     \caption{\EDIT{for old ansatz: }LCE at parameters : $n_{5}=1,~~E=12,~~r(0)=20,~~p_{r}(0)=10,~~\phi_{1}(0)=15$. The sum of the Lyapunov exponents is zero. }
%     \label{olldLLE}
% \end{figure}
% \vspace{12mm}
% \newline
%A. with $\theta=0$: see fig \ref{theta-0} \\
%B. with $\theta \ne 0$ : see fig \ref{theta-1.14}
% \vspace{12mm}
%For new ansatz, Lyapunov exponent is:
% \begin{figure}
%     \centering
%     \includegraphics[width=0.5\linewidth]{LC-new.pdf}
%     \caption{Caption}
%     \label{fig:enter-label}
% \end{figure}
%\section{Pulsating string in the D3-brane with constant NS B-field}
% \newline
% \vspace{12mm}
% \newline
The set of Lyapunov exponents for the capture scenario is 
% $\{0.275484,0.122083,-0.076478,-0.321089\}$ 
$\{{0.588666, 0.096806, -0.086377, -0.599095}\}$
whereas for an escape to infinity is 
% $\{0.000528,0.001612,0.000103,-0.002243\}$.
$\{{0.000216319, 0.277277, -0.000169858, -0.277323}\}$.
The chaos is weaker in the escape to infinity case which suggests that the presence of a horizon makes the dynamics more chaotic. Next, we shall focus on the LLE values. In Fig \ref{LLE-comparison}, we have plotted LLE for increasing value of $\theta$. It shows that our system is weakly chaotic due to small values of LLE. This also indicates that the system becomes more chaotic with the increase in the $\theta$ of the ring string for a fixed value of the energy and other initial conditions.
% When the non-commutative parameter is turned off while keeping the other parameters the same as in Fig \ref{capture-escape}(b), the system exhibits chaos since the LLE value is 0.000907 (small positive value). Therefore the chaos becomes significantly weaker in the absence of the non-commutative parameter, whereas its presence intensifies the chaotic dynamics where the LLE is 0.001612 (see Fig \ref{LLE-comparison}). Similar conclusions hold true for the capture mode as well. Thus, it is evident that the non-commutative parameter enhances chaos in the system.\\
% \begin{figure}
%     \centering
%     \includegraphics[width=0.5\linewidth]{new-2-LE-capture.pdf}
%     \caption{Comparing the LLE values for the cases when $\theta=0$ and $\theta=1.5$. The LLE is slightly larger when $\theta\ne0$ than the $\theta=0.$ The initial condition $r(0)=30,~p_{r}=0.5,~\phi_{1}(0)=0,~N=500$.}
%     \label{LLE-comparison}
% \end{figure}
% Also, we draw a table..
% \begin{table}[h]
% \centering
% \begin{tabular}{cc}
% \toprule
% $\theta$ & LLE \\
% \midrule
% Value 1 & Value 1 \\
% Value 2 & Value 2 \\
% Value 3 & Value 3 \\
% Value 4 & Value 4 \\
% \bottomrule
% \end{tabular}
% \caption{Table showing $\theta$ and corresponding LLE values.}
% \label{tab:thetaLLE}
% \end{table}

\begin{figure}
    \centering
    \includegraphics[width=0.5\linewidth]{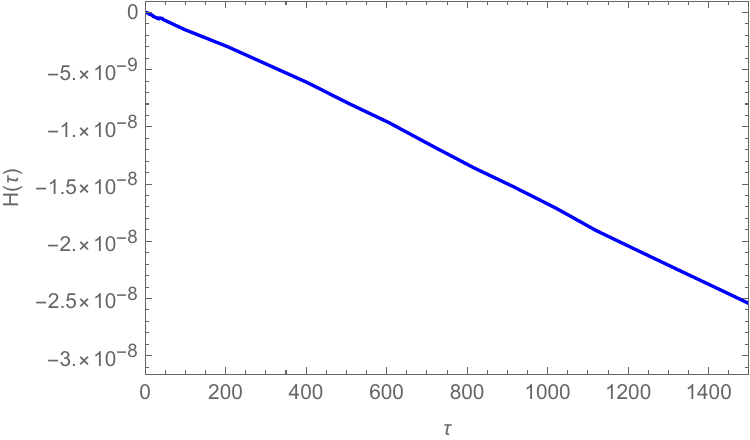}
    \caption{A typical evolution of the $H(\tau)$.}
    \label{error-plot}
\end{figure}

Before concluding this section, we would like to mention that in this paper, we employ the \texttt{Projection} method within Mathematica's \texttt{NDSolve} routine to solve the equations of motion. It is important to note that we are working with nonlinear differential equations, requiring the constraint \( H = 0 \) to be monitored and maintained at every step of the integration process. We have verified the numerical accuracy and determined that the maximum possible error, within a reasonable computation time, is on the order of \( 10^{-8} \). The Fig \ref{error-plot} shows one of the typical scenarios of the evolution of the constraint being followed. 
%some observations are the following :
%\begin{itemize}
    %\item There are two types of string trajectories: capture and escape to infinity. See fig \ref{capture-escape}.
    %\item for capture mode(see fig \ref{capture-escape} a ) the value of $\theta$ lies in the interval: $\pi/2 < \theta<3 \pi/2$ and $2\pi<\theta$
    %\item for escape(see fig \ref{capture-escape} b) to infinity : $0 < \theta< \pi/2$ and $3 \pi/2 < \theta<2 \pi$
   %\item LLE value is higher for $\theta \ne 0$
     %\item Fig \ref{theta-0} and Fig \ref{theta-1.14} both shows the chaos however in non-zero $\theta$ the chaos is stronger.
     %\item the non-commutative parameter $\theta$ causes more chaos
    % \item no oscillatory behaviour is observed and so we do not see the regular closed Poincare sections.
%\end{itemize}

% \section{Back reaction things- for Russo background}
% Maybe a similar analysis can be done ...?!!
% \\
% Can we find dispersion relation using the laplace baltrami operator ?\\
% Can we do the stability of the string in this background..??!!
% \\
% New way of non-integrability by Dibakar Roychawdhary: 
% \url{https://www.sciencedirect.com/science/article/pii/S0370269324006506}
% %\hyperlink{https://www.sciencedirect.com/science/article/pii/S0370269324006506}{}
\section{Perturbation analysis in D3-brane background}\label{perturb analysis}
% \EDIT{write some connection with the previous sections}
In this section, we shall embark on studying the D3-brane without the involvement of the NS-NS flux. We will effectively reduce the system into three dimensions and perform the perturbation analysis.
\\
The metric is given by 
\begin{align}
    ds^2 &= f^{-1/2} \Big(-dx_{0}^2 + dx_{1}^2 + h(dx_{2}^2 + dx_{3}^2) \Big) + f^{1/2} (dr^2 + r^2 d\Omega_{5}^2),
\end{align}
where, $d\Omega_{5}^2= d\phi_{1}^2 + \cos^2 \phi_{1} d\phi_{2}^2 + \sin^2 \phi_{1} \Big(d\phi_{3}^2 + \cos^2\phi_{3} d\phi_{4}^2 + \sin^2 \phi_{3} d\phi_{5}^2\Big)$.\\
% Let us consider a sting pulsating in the $S^5$ with $\phi_{1}=\phi_{3}=\frac{\pi}{2}$ with an ansatz 1 
% \begin{flalign}
%   t = t( \tau) \ ,  r= r(\tau) \ , \phi_{5} = m \sigma
% \end{flalign}
% while the rest of the coordinates are set to constant. \\
% The corresponding Polyakov action becomes
% \begin{align}
%     S_{P}=-\frac{1}{4 \pi \alpha'} \int d\tau d\sigma  (\frac{\dot t^2}{\sqrt{f}}- \sqrt{f}\dot r^2 + \sqrt{f} r^2 m^2)
% \end{align}
% The induced metric is given by,
% \begin{align}
%     ds^{2}_{ind}&= G_{\alpha \beta} d\sigma^\alpha d\sigma^{\beta}\\
%      &= m^2\sqrt{f}r^2(-d\tau^2 + d\sigma^2)
% \end{align}
% % The conserved energy is given by $\mathcal{E}=\kappa$.
% The energy is given by $\mathcal{E}=\frac{\partial \mathcal{L}}{\partial \dot t}=\frac{-\dot t}{\sqrt{f}}$.
% Then the equation of motion along r is 
% \begin{align}
%     2 \sqrt{f} \ddot{r}+\frac{f'\dot{r}^2}{2\sqrt{f}}- \frac{\dot{t}^2f'}{2f\sqrt{f}} + m^2\Big(\frac{r^2 f'}{2\sqrt{f}} +2 r\sqrt{f}\Big)=0
% \end{align}
%  Also, the Virasoro constraint or the conformal gauge constraint is given by
%  \begin{align}
%      \frac{-\dot{t}^2}{\sqrt{f}}+\sqrt{f}\dot{r}^2+\sqrt{f}r^2m^2=0
% \end{align}
% \EDIT{This ansatz is NOT consistent with the Vir constriant. }\\
% \newline
Consider the following ansatz: $\phi_{1}=\phi_{3}=\frac{\pi}{2}$ with 
\begin{flalign}
  t = t( \tau) \ ,  r= r(\sigma) \ , \phi_{5} = \phi_{5}(\tau)=\omega \tau,
\end{flalign}
while the rest of the coordinates are set to some arbitrary constant values. \\
The corresponding Polyakov action becomes
\begin{align}
    S_{P}=-\frac{1}{4 \pi \alpha'} \int d\tau d\sigma  (\frac{\dot t^2}{\sqrt{f}}- r^2 \sqrt{f} \dot \phi^2+ r'^2 \sqrt{f}).
\end{align}
The induced metric is given by,
\begin{align}
    ds^{2}_{ind}&= G_{\alpha \beta} d\sigma^\alpha d\sigma^{\beta},\\
     &=(-\frac{\dot t^2}{\sqrt{f}}+\dot \phi_{5}^2 r^2\sqrt{f})d\tau^2 + \sqrt{f} r'^2 d\sigma^2.
\end{align}
% The conserved energy is given by $\mathcal{E}=\kappa$.
The energy is given by $\mathcal{E}=\frac{\partial \mathcal{L}}{\partial \dot t}=\frac{-\dot t}{\sqrt{f}}$, whereas the equation of motion for $t$ is : $\ddot{t}=0$.\\
The equation of motion along $r$ is 
\begin{align}
    2 r'' \sqrt{f} + \frac{r'^2 f'}{2 \sqrt{f}}+ \frac{\dot t^2}{2 f^{3/2}} + 2 r \dot \phi_{5}^2+\frac{\dot\phi_{5}r^2f'}{2\sqrt{f}}=0.
\end{align}
 Also, the Virasoro constraint or the conformal gauge constraint is given by
 \begin{align}
    -\frac{\dot t^2}{\sqrt{f}}+\dot \phi_{5}^2 r^2\sqrt{f} +  \sqrt{f} r'^2 =0.
\end{align}
the above ansatz is consistent with the equation of motion (by differentiating wrt $\tau$, we get $\ddot{t}=0$ and by differentiating wrt $\sigma$, we get the $r$-equation). \\

\begin{figure}
    \centering
    \includegraphics[width=0.5\linewidth]{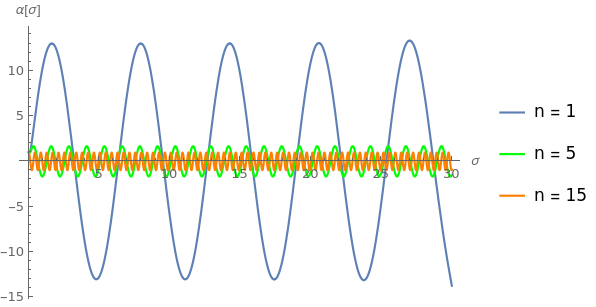}
    \caption{Plot of $\alpha$ vs $\sigma$ at $\omega=0.05,~\mathcal{E}=5,~\alpha'^2 R^4=25,~c=0$ with $n=1,5,15$.}
    \label{alpha-plot}
\end{figure}

\begin{figure}[ht!]
\begin{subfigure}
    \centering
\includegraphics[width=0.5\linewidth]{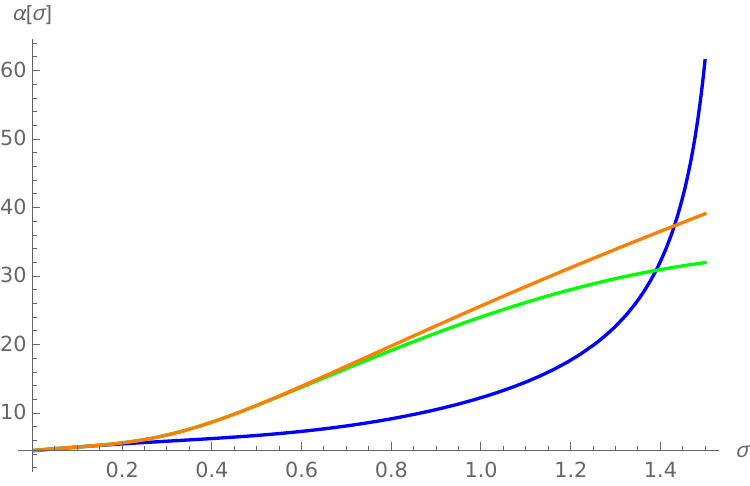}
    \put(-210,150){(a)}
    % \caption{}
    % \label{diff-plot}
\end{subfigure}
\begin{subfigure}
    \centering
\includegraphics[width=0.5\linewidth]{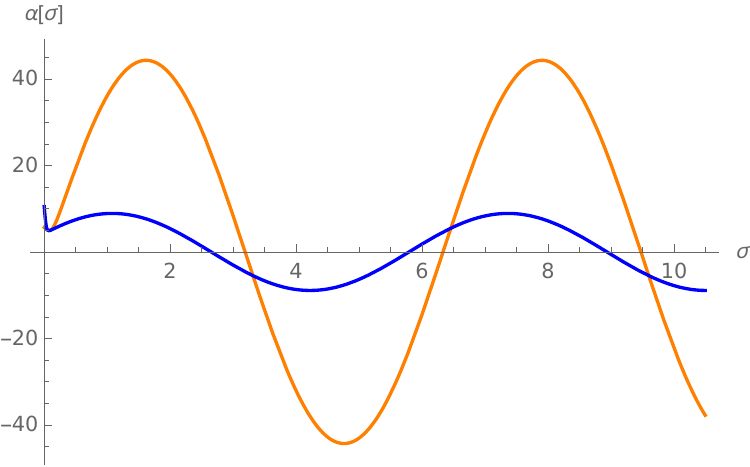}
    \put(-210,150){(b)}
    % \caption{}
    % \label{oscill-diff-E}
\end{subfigure}
\caption{Plots of $\alpha$ as a function of $\sigma$ for two scenarios: (a) $E=2,~c=0,~\alpha'=0.05,~$, $\omega=0.07\text{(green)},~0.5\text{(orange)},~1\text{(blue).}~~$ (b) $\omega=0.017,~c=0,~\alpha'=0.05,~$ $E=5(\text{oragne}), 15(\text{blue})$. In both cases, $n=1$, $R=10$.} 
\label{combined-label}
\end{figure}

Next, we consider the first-order perturbations along this string. 
% We solve the system,
%in the limit 
%where r is very large such that $O(1/r^5)\approx0$. Consequently, we have $f'\approx0$. 
The solution $r(\sigma)$ is given by 
\begin{align}
     r(\sigma) = \frac{\mathcal{E}}{\omega} \sin(\omega\sigma + C),
\end{align}
which is oscillatory.
% However, in the near horizon $r\approx0$, we have the hypergeometric solution
% \begin{align}
%     \frac{dr}{d\sigma}=\sqrt{\mathcal{E}^2- \frac{\omega^2r^6}{\alpha'^2R^4}}\\
%     \frac{r \sqrt{1-\frac{r^6 \omega ^2}{\alpha ^2 E^2 R^4}} \, _2F_1\left(\frac{1}{6},\frac{1}{2};\frac{7}{6};\frac{r^6 \omega ^2}{R^4 \alpha ^2 E^2}\right)}{\sqrt{E^2-\frac{r^6 \omega ^2}{\alpha ^2 R^4}}} = \sigma + C
% \end{align}
% we shall avoid this scenario and only consider the case when $r$ is very large such that $O(1/r^5)\approx0$.\\
Next, we determine the tangent vectors
$$\dot X=(\dot t,0,\omega),~~~~~X'=(0,r',0)$$. (where we have used here the specific form $\phi=\omega \tau$).\\
Then the normal vectors are
\begin{align}
    N^{\mu}=\Big(\frac{r^2 \omega  \sqrt[4]{f(r)}}{\sqrt{E^2 r^2-r^4 \omega ^2}},0,-\frac{E}{\sqrt[4]{f(r)} \sqrt{r^2 \left(E^2-r^2 \omega ^2\right)}} \Big).
\end{align}
The components of the extrinsic curvature tensor are given by
\begin{align}
    K_{\tau \tau}= 0,\\
     K_{\tau \sigma}=-\frac{E \sqrt[4]{f} r w \left(r f'(r)+4 f(r)\right)}{4 f(r)},\\ 
     K_{\sigma \tau}=-\frac{E \sqrt[4]{f} r w \left(r f'(r)+4 f(r)\right)}{4 f(r)},\\
      K_{\sigma \sigma}=0.
\end{align}
 By using Mathematica we found that the normal fundamental forms vanish. Next, we find the first-order perturbations equation from 
\begin{equation}
\Box \varphi^i + 2 \mu^\alpha_{ij} \partial_\alpha \varphi^j + (\nabla_\alpha \mu^\alpha_{ij}) \varphi^j - \mu^\alpha_{il} \mu^l_{j\alpha} \varphi^j 
+ \frac{2}{G_{cc}} K^{\alpha\beta}_i K_{\alpha\beta}^j \varphi^j - h^{\alpha\beta} R^\mu_{\rho\sigma\nu} N^\rho_i N^\sigma_j X^\mu_{,\alpha} X^\nu_{,\beta} \varphi^j = 0.
\end{equation}
Computing the term $\frac{2}{G_{cc}} K^{\alpha\beta}_i K_{\alpha\beta}^j$ which is given by
\begin{align}
    \frac{2}{G^{c}_{c}} K^{\alpha\beta}_i K_{\alpha\beta}^j =-\frac{E^2 \omega ^2 \left(r f'(r)+4 f(r)\right)^2}{8 f(r)^2 (E-r \omega ) (E+r \omega )},
\end{align}
and the term 
\begin{align}
h^{\alpha\beta} R^\mu_{\rho\sigma\nu} N^\rho_i N^\sigma_j X^\mu_{,\alpha} X^\nu_{,\beta} \varphi^j=\frac{4 f(r) \left(E^2 f''(r)+r \omega ^2 \left(r f''(r)+f'(r)\right)\right)-5 f'(r)^2 \left(E^2+r^2 \omega ^2\right)}{16 f(r)^2}.
\end{align}
% \EDIT{(All these are found by using Mathematica.)\\}
Thus, we get the first-order perturbations equation as 
\begin{align}
    -\ddot \varphi + \varphi'' -\frac{E^2 \omega ^2 \left(r f'(r)+4 f(r)\right)^2}{8 f(r)^2 (E-r \omega ) (E+r \omega )}\varphi \nonumber \\
    -\frac{4 f(r) \left(E^2 f''(r)+r \omega ^2 \left(r f''(r)+f'(r)\right)\right)-5 f'(r)^2 \left(E^2+r^2 \omega ^2\right)}{16 f(r)^2} \varphi=0.
    \label{pert-eq}
\end{align}
Now, we can take the Fourier series expansion to be $\varphi=\sum_{n} \alpha_{n}(\sigma)e^{i n \tau}$, we get
\begin{align}
   \alpha ''(\sigma )+\alpha (\sigma ) \Big(n^2-\frac{4 f(r) \left(E^2 f''(r)+r \omega ^2 \left(f'(r)+r f''(r)\right)\right)-5 f'(r)^2 \left(E^2+r^2 \omega ^2\right)}{16 f(r)^2} \nonumber \\  
   -\frac{E^2 \omega ^2 \left(r f'(r)+4 f(r)\right)^2}{8 f(r)^2 (E-r \omega ) (E+r \omega )}\Big)=0.
\end{align}
 Next, we numerically solve this equation using solution  $r(\sigma) = \frac{\mathcal{E}}{\omega} \sin(\omega\sigma + C)$. 
We plot $\alpha-\sigma$ in Fig. \ref{alpha-plot}. From the plot, it is clear that the amplitude decreases as $n$ increases. 
In the limit $\mathcal{E}$ is large, the equation reduces significantly to the well known Poschl-Teller equation :
\begin{align}
\alpha''(\sigma) + \left[n^2 - 2\omega^2 \sec^2(c + \sigma \omega)\right] \alpha(\sigma) = 0.
\end{align}
The solutions are expressed in terms of associated Legendre functions or hypergeometric functions
\begin{align}
\alpha(\sigma) = C_1 P_{\nu}^{\mu}(\tan(c + \sigma \omega)) + C_2 Q_{\nu}^{\mu}(\tan(c + \sigma \omega)).
 \end{align}
 Except for the few special cases discussed in the section, solving the equations analytically is futile. The proper approach is to solve them numerically.
 The problem can be viewed as solving the second-order differential equation
 \begin{align}
 \alpha''(\sigma)+ G(\sigma)\alpha(\sigma)=0,~~\alpha(a_{0})=a,~~\alpha'(a_{0})=b,
 \end{align}
 where,
 \begin{align}
     G(\sigma)=n^2-\frac{2 E^8 \omega ^2 \sec ^2(c+\sigma  \omega )}{\left(\text{$\alpha'$}^2 R^2 \omega ^4 \csc ^4(c+\sigma  \omega )+E^4\right)^2} \nonumber \\ +\frac{\text{$\alpha'$}^2 R^2 \omega ^6 \csc ^4(c+\sigma  \omega ) \left(-5 E^4 \csc ^2(c+\sigma  \omega ) +\text{$\alpha'$}^2 R^2 \omega ^4 \csc ^4(c+\sigma  \omega )-4 E^4\right)}{\left(\text{$\alpha'$}^2 R^2 \omega ^4 \csc ^4(c+\sigma  \omega )+E^4\right)^2},
 \end{align}
 where $a_{0},~a,~b$ are reals.
 Ideally, one would test the stability of the equations for an infinite range of initial conditions. However, as this is impractical, we have restricted our analysis to a finite set of initial conditions. We now present the numerical results
\begin{itemize}
     % \item The effect of $n$ is to decrease the amplitude of the oscillation. See Fig \ref{alpha-plot}.
     \item The parameter $n$ plays a crucial role in modulating the dynamics of the system by reducing the amplitude of oscillations. As $n$ increases, the system experiences a damping-like effect, leading to smaller oscillatory motion. See Fig. \ref{alpha-plot}.
    % \item For oscillaotry/non-oscillatory solutions, the value of $\omega$ and $\mathcal{E}$ are also affected by the initial conditions. 
    \item For both oscillatory and non-oscillatory solutions, the values of the frequency parameter $\omega$ and the energy parameter $\mathcal{E}$ are significantly influenced by the initial conditions. 
    % \item As $\omega$ increases, $\mathcal{E}$ must decrease for a fixed set of initial conditions to ensure an oscillatory solution.
    \item For a fixed set of initial conditions, as the frequency parameter $\omega$ increases, the energy parameter $\mathcal{E}$ must decrease to maintain an oscillatory solution. This relationship ensures that the system remains within the regime where oscillations are sustained.
    % \item The solution becomes unstable for some of the values of the parameters see Fig \ref{combined-label} a.
    \item The solution becomes unstable for some of the parameter values, as illustrated in Fig.~\ref{combined-label} (a). This instability arises when the system's dynamics are sensitive to specific parameter choices, leading to a breakdown in the oscillatory.
    % \item When $\omega$ is very small, the value of $\mathcal{E}$ and the initial conditions have no significant impact, the solution still remains oscillatory. (See Fig \ref{combined-label} b). 
    \item When $\omega$ is very small, the value of $\mathcal{E}$ and the initial conditions have no significant impact, and the solution still remains oscillatory. This insensitivity to $\mathcal{E}$ and initial conditions highlights the robustness of the oscillatory behavior in the small $\omega$ limit. (See Fig.~\ref{combined-label} (b)).
\end{itemize}
\section{Pulsating string in deformed $(AdS_{3}\times S^{2})_{\varkappa}$} \label{short-long}
This section will consider the semiclassical quantization of a closed pulsating string in the deformed $(AdS_{3}\times S^{2})_{\varkappa}$ subsector of the deformed $(AdS_{5}\times S^{5})_{\varkappa}$.
  The metric of deformed $(AdS_{3} \times S^2)_{\varkappa}$ is given by \cite{hoare2014deformations,panigrahi2014pulsating}
  \begin{equation}
      ds^2_{({AdS_{3} \times S^{2}})_{\varkappa}}= - h(\rho) dt^2 + f(\rho) d\rho^2 + \rho^2 d\psi_{1}^2 + \Tilde{h}(r) d\phi^2 + \Tilde{f}(r) dr^2,
  \end{equation}
  where,
  \begin{eqnarray}
      h= \frac{1+ \rho^2}{1- \varkappa^2 \rho^2}, ~~~ f= \frac{1}{(1+ \rho^2)(1- \varkappa^2 \rho^2)}, \\
      \Tilde{h} =\frac{1-r^2}{1+ \varkappa^2 r^2}, ~~~\Tilde{f}=\frac{1}{(1-r^2)(1+\varkappa^2 r^2)},
  \end{eqnarray}
  where the coordinates $t,~\rho,~\psi_{1}$ are the coordinates of the deformed metric $AdS_{3}$ whereas the coordinates $\phi, ~r$ belong to the deformed sphere $S^{2}$.
  We do the following substitution in the metric: 
  \begin{equation}
      \rho=\sinh\rho,~~r=\cos\psi,
  \end{equation}
  then we obtain
  \begin{align}
      ds^2=& -\frac{\cosh^2 \rho}{1- \varkappa^2 \sinh^2 \rho} dt^2 + \frac{1}{1- \varkappa^2 \sinh^2\rho} d\rho^2 + \sinh^2\rho d\psi_{1}^2 + \frac{\sin^2\psi}{1+\varkappa^2 \cos^2\psi} d\phi^2 \nonumber\\
      &\quad + \frac{1}{1+\varkappa^2 \cos^2\psi} d\psi^2.
  \end{align}
  We rename the coordinate $\psi_{1} \equiv \varphi$ for notational conveniences.
 Consider the following pulsating string ansatz: 
  \begin{align}
      t=t(\tau), ~~\rho=\rho(\tau),~~\phi=\phi(\tau),~~\psi=\psi(\sigma).
  \end{align}
  % \EDIT{This ansatz doesn't include the B-field in the action. Can we do similar analysis for B-field ??!}\\
  Then, the Polyakov action becomes 
  \begin{align}
      S_{P}&=\frac{1}{4 \pi \alpha'}\sqrt{1+ \varkappa^2}\int d\tau d\sigma \Bigg(-\frac{\cosh^2 \rho}{1- \varkappa^2 \sinh^2 \rho} \dot{t}^2 + \frac{1}{1- \varkappa^2 \sinh^2\rho} \dot\rho^2  \nonumber\\
      &\quad + \frac{\sin^2\psi}{1+\varkappa^2 \cos^2\psi}\dot{\phi}^2 - \frac{1}{1+\varkappa^2 \cos^2\psi} \psi^{'2} \Bigg).
  \end{align}
  %where, $\hat{\lambda}= \lambda(1+\chi^2)$ where $\lambda$ is the 't Hooft coupling.\\
% \EDIT{This reduction is very close to what is done in ref \cite{barik2018spinning} where they had the additional $\sinh^2\rho$. \\}
%  Further, maybe due to the gauge choice they had the different signs in the action(??).}
  Therefore the Lagrangian is given by 
  \begin{align}
     \mathcal{L}&= \frac{1}{4 \pi} \Bigg(-\frac{\cosh^2 \rho}{1- \varkappa^2 \sinh^2 \rho} \dot{t}^2 + \frac{1}{1- \varkappa^2 \sinh^2\rho} \dot\rho^2  \nonumber\\
      &\quad + \frac{\sin^2\psi}{1+\varkappa^2 \cos^2\psi}\dot{\phi}^2 - \frac{1}{1+\varkappa^2 \cos^2\psi} \psi^{'2} \Bigg).
  \end{align}
  The equation of motion for t is :
  \begin{align}
     \cosh^2\rho ~\ddot t + \frac{\left(1+ \varkappa ^2\right) \sinh 2 \rho }{1-\varkappa ^2 \sinh ^2\rho} \dot t \dot \rho =0.
  \end{align}
  The equation of motion for $\rho$ is given by
   \begin{align}
  %      2 \ddot \rho (1 - \chi^2 \sinh^2 \rho) + \sinh2\rho [\chi^2 \dot\rho^2 + (1+\chi^2)\dot t^2 ]=0
  \ddot \rho \left( 2+ \varkappa ^2  -\varkappa ^2 \cosh2 \rho \right)+\sinh2 \rho \left(\varkappa ^2 \dot \rho ^2+\left(\varkappa ^2+1\right) \dot t^2\right)=0.
   \end{align}
 The equation for $\phi$ is
 \begin{align}
     \dot\phi=\omega ~~\textit{(constant)},
 \end{align}
  and the equation for $\psi$ is
  \begin{align}
     %2(1+ \chi^2 \cos^2\psi) \psi^{''2} + \psi^{'2}\chi^2 \sin2\theta -\omega^2(1+\chi^2)\sin 2\theta =0
     \sin 2 \psi \left(w^2 \left(\varkappa ^2+1\right)-\varkappa ^2 \psi '\right)^2-\psi '' \left(\varkappa ^2 \cos 2 \psi +\varkappa ^2+2\right)=0.
  \end{align}
%  And finally the equation for $\phi$ is
 % \begin{align}
     % \sin2\psi [- (1+ \chi^2)\dot \phi^2 + \chi^2 \dot\psi^2] + 2 (1+ \chi^2 \cos^2\psi) \ddot\psi =0
%\end{align}
The first Virasoro constraint is trivially satisfied, whereas the second Virasaro constraint is given by
\begin{equation}\label{vir}
    -\frac{\cosh^2 \rho}{1- \varkappa^2 \sinh^2 \rho} \dot{t}^2 + \frac{1}{1- \varkappa^2 \sinh^2\rho} \dot\rho^2  + \frac{\sin^2\psi}{1+\varkappa^2 \cos^2\psi}\dot{\phi}^2 + \frac{1}{1+\varkappa^2 \cos^2\psi} \psi^{'2} =0.
\end{equation}
% \EDIT{The ansatz 2 is consistent with the Virasoro constraint.\\ }\\
From the equation of motion of $t$:
\begin{align}
    \dot t= \frac{A (1 -\varkappa^2 \sinh^2\rho)}{\cosh^2\rho},
\end{align}
where A is a constant. 
Next, we substitute the value of $\dot t$ and $\dot \phi$ into the Virasoro constraint \ref{vir} to obtain 
\begin{align}
 -\frac{A^2 (1- \varkappa^2 \sinh^2 \rho)}{\cosh^2 \rho} + \frac{ \dot\rho^2}{1- \varkappa^2 \sinh^2\rho}  + \frac{ w^2  \sin^2\psi}{1+\varkappa^2 \cos^2\psi}+ \frac{1}{1+\varkappa^2 \cos^2\psi} \psi^{'2} =0.
\end{align}
which is a coupled differential equation in variables $\rho(\tau)$ and $\psi(\sigma)$. To solve this coupled equation, we split the equation into the variables as follows  
\begin{align}
    -\frac{A^2 (1- \varkappa^2 \sinh^2 \rho)}{\cosh^2 \rho} + \frac{ \dot\rho^2}{1- \varkappa^2 \sinh^2\rho} + C^2 &=0,\\
    \frac{ w^2 \sin^2\psi}{1+\varkappa^2 \cos^2\psi} + \frac{1}{1+\varkappa^2 \cos^2\psi} \psi^{'2}- C^2 &=0,
\end{align}
where $C$ is an arbitrary constant. 
On solving the above equations, we obtain
\begin{align}
    \sinh \rho = \sqrt{\frac{R_{-} sn^2\left(2 C \varkappa \sqrt{R_{+}} \tau, \frac{R_{-}}{R_{+}} \right)}{1+\varkappa^2 \left[1+ R_{-}sn^2\left(2 C \varkappa \sqrt{R_{+}} \tau, \frac{R_{-}}{R_{+}} \right) \right]}}, \\
    \sin\theta=\sqrt{\frac{C^2 (1+\varkappa^2)}{w^2+\varkappa^2 C^2}}sn\left(\sqrt{w^2+\varkappa^2 C^2}\sigma, \frac{C^2 (1+\varkappa^2)}{w^2+\varkappa^2 C^2} \right),
\end{align}
where,
\begin{align}
    R_{\mp}=\frac{-3C^2-A^2\varkappa^2 \mp \sqrt{A^4 \varkappa^4 + 2 A^2 C^2\varkappa^2(5+2\varkappa^2)+C^4(9+4 \varkappa^2 -4 \varkappa^4)}}{2 C^2 \varkappa^2}.
\end{align}
The conserved charges associated to the string can be given by
\begin{align}
    E=-\int^{2 \pi}_{0} \frac{\partial \mathcal{L}}{\partial \dot t}
    = A \sqrt{1+\varkappa^2}/\alpha' ,\\
    \mathcal{J_{\phi}}= \int^{2 \pi}_{0} \frac{\partial \mathcal{L}}{\partial \dot \phi} 
    =  \frac{\sqrt{1+\varkappa^2}}{2\pi \alpha'}\frac{w  \sin^2 \psi}{1+ \varkappa^2 \cos^2\psi} .\label{J-eq}
\end{align}
We define, 
\begin{align}
    \alpha \equiv \sinh \xi_{\textit{max}} = \sqrt{R_{-}},\\
    \beta \equiv \sin \theta_{\textit{max}} = \sqrt{\frac{C^2 (1+\varkappa^2)}{w^2 + \varkappa^2 C^2}}.
\end{align}
From the periodicity condition of $\theta$, we obtain
\begin{align}
    2 \pi \sqrt{w^2 + \varkappa^2 C^2} = 4 \textbf{K}\left[\frac{C^2(1+\varkappa^2)}{w^2 + \varkappa^2 C^2}\right].
\end{align}
where \textbf{K}(..) is complete elliptic integral of first kind.
Thus, consequently, we get
\begin{align}
    C=\frac{2 \beta \textbf{K}(\beta^2)}{\pi \sqrt{1+\varkappa^2}}, ~~~ w= \frac{2}{\pi} \sqrt{\frac{1+\varkappa^2 (1- \beta^2)}{1+\varkappa^2}} \textbf{K}(\beta^2). \label{C-w}
\end{align}
Next, substituting the value of $\sin\theta$ into eq \ref{J-eq}, we get
\begin{align}
    \mathcal{J}_{\phi} = \frac{2\sqrt{1+\varkappa^2-\varkappa^2 \beta^2}}{\pi \alpha' \varkappa^2} \Big[\Pi\Big(\frac{\varkappa^2 \beta^2}{1+\varkappa^2},\beta^2\Big) -\textbf{K}(\beta^2)\Big].
\end{align}
For solving the expression of $E$, we first find $A$, then substitute the value of $C$ from eq \ref{C-w}, thus we get the conserved charge as follows
\begin{align}
    E=\frac{8 \beta  \textbf{K}\left(\beta ^2\right) \sqrt{ 1+\varkappa ^2} \cosh \xi}{\alpha'  \sqrt{ 1+\varkappa ^2}},\\
    =\frac{8}{\alpha'}\beta \sqrt{1+\alpha ^2} \textbf{K}\left(\beta ^2\right).
\end{align}
In the next section, we discuss and evaluate the expressions of the energy and angular momentum for short and long strings.
\subsection{Short string limit}
Let us now consider a short string solution in $S^2$, therefore, $\sin\theta \ll 1$ or, $\beta \ll 1$. On applying, the energy becomes 
\begin{align}
    E \approx \frac{4 \pi}{\alpha'} \beta \sqrt{1+\alpha^2},
\end{align}
and the angular momentum 
\begin{align}
    \mathcal{J}_{\phi} \approx \frac{\beta^2}{2 \alpha'^2 \sqrt{1+\varkappa^2}},
\end{align}
thus, from these two expressions, we get the dispersion relation as follows
\begin{align}
    E^2 \approx 32 \pi^2 \sqrt{1+
    \varkappa^2} \sqrt{1+\alpha^2}\mathcal{J}_{\phi}.
\end{align}
When the string is at the centre of $AdS_{3}$ and performs the small oscillation then $\alpha<<1$, thus
\begin{align}
    E^2 \approx 32 \pi^2 \sqrt{1+
    \varkappa^2} \mathcal{J}_{\phi},
\end{align}
and in the limit $\varkappa \xrightarrow{}0$ (undeformed case of $AdS_{3} \times S^2$), we get $E^2 \approx 32 \pi^2 \mathcal{J}_{\phi}$. Further, when $\alpha>>1$, we write the dispersion relation as follows
\begin{align}
    E\approx \pi  \sqrt{2} \sqrt{\mathcal{J}_{\phi}}+\pi  \sqrt{2} \sqrt{\mathcal{J}_{\phi}} \varkappa ^2+\frac{\pi  \alpha ^2 \sqrt{\mathcal{J}_{\phi}} \varkappa ^2}{2 \sqrt{2}}+\frac{\pi  \alpha ^2 \sqrt{\mathcal{J}_{\phi}}}{2 \sqrt{2}}.
    %-\frac{3 \pi  \alpha ^4 \sqrt{J} \chi ^2}{16 \sqrt{2}}-\frac{3 \pi  \alpha ^4 \sqrt{J}}{16 \sqrt{2}}
\end{align}
In this case, the undeformed case leads to $E\approx \pi  \sqrt{2} \sqrt{\mathcal{J}_{\phi}} + \frac{\pi  \alpha ^2 \sqrt{\mathcal{J}_{\phi}}}{2 \sqrt{2}} $.
\subsection{Long string limit}
In this section, we take the limit $\beta \approx 1$ which corresponds to the long string case extending to the equator of the $S^2$. Then the energy becomes as 
\begin{align}
    E\approx \frac{4}{\alpha'} \sqrt{1+\alpha^2} \textit{ln}\frac{16}{1- \beta^2},
\end{align}
and the angular momentum is 
\begin{align}
    \mathcal{J}_{\phi} \approx \frac{1}{\alpha' \pi} ln\frac{16}{1-\beta^2} - \frac{6+2\varkappa^2}{3 \pi \alpha'},
\end{align}
so the relation between $E$ and $\mathcal{J}_{\phi}$ is
\begin{align}
    E \approx \frac{4}{\alpha'} \sqrt{1+\alpha^2}(\mathcal{J}_{\phi} \alpha' \pi + 2+ \frac{2 \varkappa^2}{3}).
\end{align}
In the small oscillation ie $\alpha << 1$, 
\begin{align}
    E\approx 4 \pi \mathcal{J}_{\phi} + \frac{8}{\alpha'} +\frac{8 \varkappa^2}{3 \alpha'},
\end{align}
whereas in the limit $\alpha>>1$, we get
\begin{align}
    E\approx \frac{4}{\alpha'}\alpha ~\textit{ln}\frac{16}{1-\beta^2}.
\end{align}
\section{Conclusions and Future directions} \label{conclusion}
% \EDIT{We can do a similar analysis(non-integrability, chaos and dispersion relations etc) on this deformed background as well.\\
% Can be extended to the involvement of the B-field\\
% }\\
% \newline
% We begin by establishing first the non-integrability and chaos in the D3-brane with the effective non-commutative parameter. Further, we also discuss the point-particle case in which the dynamics have been proven to be integrable. It is quite interesting that the non-commutative parameter enhances the effect of the chaos in the system. Also, we observe two different string modes namely capture and escape to infinity. The Lyapunov values are slightly higher in the capture case than in the escape to infinity. 
In this work, we start by investigating the (non)-integrability and chaotic behaviour of the D3-brane in the presence of an effective non-commutative parameter. Additionally, we examine the point-particle case, where the dynamics are shown to be integrable. Notably, the non-commutative parameter intensifies the chaotic effects within the system. Moreover, we identify two distinct string modes: one leading to captured at and the other escaping to infinity. The (largest) Lyapunov exponents are found to be slightly higher in the capture scenario compared to the escape-to-infinity case. \\
% By using the general formalism of the construction of the perturbation equation via the Polyakov action, the geometric covariant quantities like normal fundamental form and extrinsic curvature tensor have been computed first to write the perturbation equations. Then using the pulsating string ansatz, we derive solutions to the equations of motion and constraints in the given background. In the large \(E\) limit, the first-order perturbed equation, expressed in Fourier series form, simplifies to the Pöschl-Teller equation, whose solutions are given in terms of associated Legendre functions or hypergeometric functions. For generic values of \(E\), the corresponding second-order equation is solved numerically. This study is expected to aid in determining the first-order correction to the energy, which corresponds to the anomalous dimension of gauge theory operators in the strong coupling regime.\\
Next, using the Polyakov action, with the help of geometric quantities like the extrinsic curvature tensor, we formulate perturbations equations. With a pulsating string ansatz, solutions to the motion and constraints were derived. In the high-energy limit, the first-order equation simplifies to the Pöschl-Teller equation, solvable via Legendre or hypergeometric functions, while numerical methods address generic \(\mathcal{E}\) cases. This aids in finding first-order energy corrections linked to anomalous dimensions in strongly-coupled gauge theory. Moreover, the numerical results show that for oscillatory/non-oscillatory solutions, \(\omega\) and \(\mathcal{E}\) depend on initial conditions. As \(\omega\) increases, \(\mathcal{E}\) must decrease to maintain oscillations. Small \(\omega\) ensures oscillatory solutions regardless of \(\mathcal{E}\) or initial conditions, while certain parameter values lead to instability.\\
Lastly, we have studied pulsating string in the so-called deformed $(AdS_3\times S^2)_{\varkappa}$ background. To the best of our knowledge, the analysis of the string dispersion relation in the undeformed case has not been done before. To obtain pulsating string solutions in the deformed background, we have employed a suitable ansatz for a circular pulsating string configuration. By considering the bosonic sector of the string action in the conformal gauge, we derived the corresponding equations of motion and Virasoro constraints. We have further found out the short string energy as a function of $\mathcal{J}_{\phi}$ and $\varkappa$, and for the long string, in the limit of $\alpha>>1$ it is an involved logarithmic function of $\varkappa$. Thus, the energy and the computed corrections serve as the first examples of contributions from this string sector to the anomalous dimensions in the dual gauge theory. In future, it would be interesting to include the effect of the B-field using a different string ansatz and analyze its influence on chaos in the system, as this could reveal new dynamical features. Further, it is crucial to investigate field theory duals of these theories, which would provide a deeper understanding of the gauge-gravity correspondence and its implications.
\newline

 \bibliographystyle{JHEP}
\bibliography{pulsating.bib}
  % \bibliography{biblio.bib}

%% or
%% [B] Manual formatting (see below)
%% (i) We suggest to always provide author, title and journal data or doi:
%% in short all the informations that clearly identify a document.
%% (ii) please avoid comments such as "For a review'', "For some examples",
%% "and references therein" or move them in the text. In general, please leave only references in the bibliography and move all
%% accessory text in footnotes.
%% (iii) Also, please have only one work for each \bibitem.

% \begin{thebibliography}{99}

% \bibitem{a}
% Author,
% \emph{Title},
% \emph{J. Abbrev.} {\bf vol} (year) pg.

% \bibitem{b}
% Author,
% \emph{Title},
% arxiv:1234.5678.

% \bibitem{c}
% Author,
% \emph{Title},
% Publisher (year).

% \end{thebibliography}
\end{document}